\documentclass[%
 reprint,
superscriptaddress,
 amsmath,amssymb,
prd,
]{revtex4-2}

\usepackage{graphicx}
\usepackage{dcolumn}
\usepackage{aas_macros}
\usepackage{bm}
\usepackage{xcolor}
\usepackage{comment}
\definecolor{turquoise}{RGB}{96, 158, 160}
\definecolor{fuchsia}{RGB}{255, 0, 255}
\definecolor{darkorange}{RGB}{255, 127, 79}
\definecolor{red}{RGB}{220, 20, 60}
\definecolor{green}{RGB}{4, 235, 2}
\definecolor{darkyellow}{RGB}{255, 166, 6}

\def\PN{{\cal PN}}

\usepackage{hyperref}
\usepackage[mathlines]{lineno}

\definecolor{seagreen}{rgb}{0.190, 0.525, 0.361}

\begin{document}

\preprint{APS/123-QED}

\title{Large Gravitational Wave Phase Shifts \\ from Strong 3-body Interactions in Dense Stellar Clusters}

\author{Kai Hendriks}
\thanks{Corresponding author: kai.hendriks@nbi.ku.dk}
\affiliation{Niels Bohr International Academy, The Niels Bohr Institute, Blegdamsvej 17, DK-2100, Copenhagen, Denmark}

\affiliation{Department of Physics \& Astronomy, Northwestern University, Evanston IL 60208, USA}
\affiliation{Center for Interdisciplinary Exploration \& Research in Astrophysics (CIERA), Evanston, IL}

\author{Dany Atallah}
\affiliation{Department of Physics \& Astronomy, Northwestern University, Evanston IL 60208, USA}
\affiliation{Center for Interdisciplinary Exploration \& Research in Astrophysics (CIERA), Evanston, IL}

\author{Miguel Martinez}
\affiliation{Department of Physics \& Astronomy, Northwestern University, Evanston IL 60208, USA}
\affiliation{Center for Interdisciplinary Exploration \& Research in Astrophysics (CIERA), Evanston, IL}

\author{Michael Zevin}
\affiliation{Adler Planetarium, 1300 South DuSable Lake Shore Drive, Chicago, IL, 60605, USA}
\affiliation{Center for Interdisciplinary Exploration \& Research in Astrophysics (CIERA), Evanston, IL}

\author{Lorenz Zwick}
\affiliation{Niels Bohr International Academy, The Niels Bohr Institute, Blegdamsvej 17, DK-2100, Copenhagen, Denmark}

\author{Alessandro A. Trani}
\affiliation{Niels Bohr International Academy, The Niels Bohr Institute, Blegdamsvej 17, DK-2100, Copenhagen, Denmark}
\affiliation{Department of Astronomy, University of Concepti\'on, Avenida Esteban Iturra s/n Casilla 160-C Concepti\'on, Chile}
\affiliation{INFN, Sezione di Trieste, I-34127, Trieste, Italy}

\author{Pankaj Saini}
\affiliation{Niels Bohr International Academy, The Niels Bohr Institute, Blegdamsvej 17, DK-2100, Copenhagen, Denmark}

\author{J\'{a}nos Tak\'{a}tsy}
\affiliation{Niels Bohr International Academy, The Niels Bohr Institute, Blegdamsvej 17, DK-2100, Copenhagen, Denmark}

\author{Johan Samsing}
\affiliation{Niels Bohr International Academy, The Niels Bohr Institute, Blegdamsvej 17, DK-2100, Copenhagen, Denmark}

\date{\today}

\begin{abstract}

The phase evolution of gravitational waves (GWs) can be modulated by the astrophysical environment surrounding the source, which provides a probe for the origin of individual binary black holes (BBHs) using GWs alone. We here study the evolving phase of the
GW waveform derived from a large set of simulations of BBH mergers forming in dense stellar clusters through binary-single interactions.
We uncover that a well-defined fraction of the assembled eccentric GW sources will have a notable GW phase shift induced
by the remaining third object. The magnitude of the GW phase shift often exceeds conservative analytical estimates due to strong 3-body interactions, which occasionally results in GW sources with clearly shifted and perturbed GW waveforms.
This opens up promising opportunities for current and future GW detectors, as observing such a phase shift can identify the formation environment of a BBH, as well as help to characterise the local properties of its surrounding environment.

\end{abstract}

\maketitle

\section{Introduction}\label{sec:Introduction}

Gravitational wave (GW) templates for binary black hole (BBH) mergers are constructed with the masses, spins, and orbital eccentricities of the progenitor systems. The different proposed formation channels of BBH produce distinct features in these three parameters, indicative of the BBH formation mechanism \citep[e.g.][]{2017ApJ...846...82Z,2021MNRAS.505.3681S, 2006ApJ...640..156G, 2014ApJ...784...71S, 2017ApJ...840L..14S, Samsing18a, Samsing2018, 2018ApJ...855..124S,
2018MNRAS.tmp.2223S, 2018PhRvD..98l3005R, 2019ApJ...881...41L}. While powerful, this method only allows for population-level statements and cannot on its own directly show how a \textit{single} BBH formed.

A novel approach that is gaining increasing attention is to identify direct modulations of the GW signals of binaries due to the astrophysical environment in which they form, which, if observable, yield a means to probe the environment and formation
mechanism of the BBH itself on a \textit{single-event basis} \citep[e.g.][]{2023ApJ...954..105V}. These modulations include general relativistic propagation effects (e.g., GW source acceleration \citep[e.g.][]{2011PhRvD..83d4030Y, 2017ApJ...834..200M, 2017PhRvD..96f3014I, 2018PhRvD..98f4012R, 2019PhRvD..99b4025C,
2021PhRvL.126b1101Y, 2019ApJ...878...75R, 
2019MNRAS.488.5665W, 2020PhRvD.101f3002T, 2020PhRvD.101h3031D, 2021PhRvL.126j1105T, 
2021PhRvD.104j3011Y, 2022PhRvD.105l4048S, 2023PhRvD.107d3009X, 2023arXiv231016799L, 2023ApJ...954..105V, 2024arXiv240305625S, 2024arXiv240804603H}, GW lensing and gravitational redshift \citep[e.g.][]{2022MNRAS.515.3299G, 2024PhRvD.110d4054P}), astrophysical environmental effects (e.g., gas dynamical friction \citep[e.g.][]{2014barausse, 2023MNRAS.521.4645Z} and tidal forces \citep{2024arXiv240305625S}), and effects beyond classical general relativity (GR) \citep[e.g.][]{2023PhRvD.107h4011C, 2023arXiv231006894C}. As no BBH merges in a completely empty Universe, such effects should leave imprints in every observed GW signal;
however, in most cases the modulations are too small to be resolved. This has led to the key question whether any of the considered
BBH merger formation channels such as stellar clusters \citep{2000ApJ...528L..17P, Lee:2010in,
2010MNRAS.402..371B, 2013MNRAS.435.1358T, 2014MNRAS.440.2714B,
2015PhRvL.115e1101R, 2015ApJ...802L..22R, 2016PhRvD..93h4029R, 2016ApJ...824L...8R,
2016ApJ...824L...8R, 2017MNRAS.464L..36A, 2017MNRAS.469.4665P, 2018PhRvD..98l3005R, 2018PhRvD..97j3014S,
2018MNRAS.tmp.2223S, 2020PhRvD.101l3010S, 2021MNRAS.504..910T, 2021MNRAS.504..910T, 2022MNRAS.511.1362T, 2024A&A...689A..24T},
isolated binary stars \citep{2012ApJ...759...52D, 2013ApJ...779...72D, 2015ApJ...806..263D, 2016ApJ...819..108B,
2016Natur.534..512B, 2017ApJ...836...39S, 2017ApJ...845..173M, 2018ApJ...863....7R, 2018ApJ...862L...3S, 2019MNRAS.485..889S, 2021ApJ...910...30T, 2022ApJ...926...83T, 2023MNRAS.524..426I},
active galactic nuclei (AGN) discs \citep{2017ApJ...835..165B,  2017MNRAS.464..946S, 2017arXiv170207818M, 2020ApJ...898...25T,
2022Natur.603..237S, 2023arXiv231213281T, Fabj24}, or galactic nuclei \citep{2009MNRAS.395.2127O, 2015MNRAS.448..754H,
2016ApJ...828...77V, 2016ApJ...831..187A, 2016MNRAS.460.3494S, 2017arXiv170609896H, 2018ApJ...865....2H,
2019ApJ...885..135T, 2019ApJ...883L...7L,2021MNRAS.502.2049L, 2023MNRAS.523.4227A, 2024A&A...683A.135T},
naturally produce a notable fraction of mergers with measurable GW modulations able to inform about the environment of the merger.

\begin{figure*}
    \centering
    \includegraphics[width=1.\textwidth]{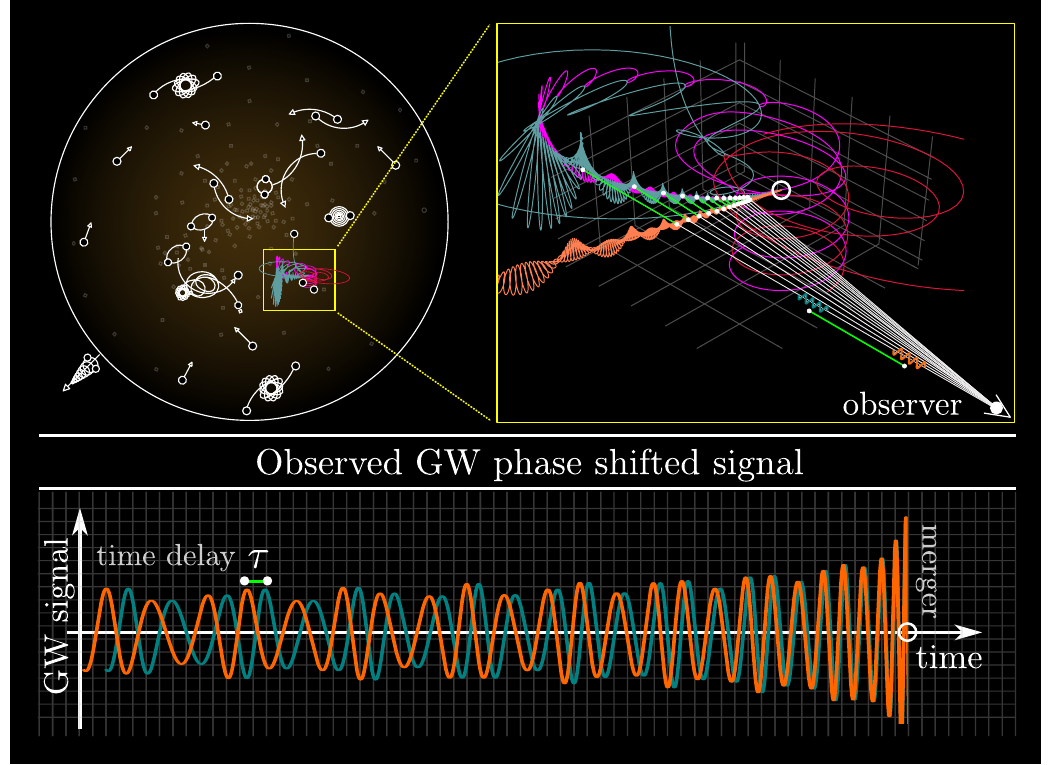}
    \caption{{\bf Illustration of 3-body interaction resulting in a BBH merger with an observable GW Phase Shift.}
    \textit{Top left}: A stellar cluster with highlighted BH interactions, each of which is able to produce BBH mergers. \textit{Top right}: Zoom-in
    on a binary-single interaction resulting in a BBH merger with the third object still bound (3-body merger). {\it Turquoise-} and {\it pink}
    lines show the trajectory of the merging BHs (true path), where the {\it orange} lines illustrate the path the BBH would have taken without
    the third object (reference path). The trajectory of the third BH is depicted in {\it red}. The {\it white} lines show lines-of-sight for an
    observer located in the lower right corner, where the {\it green} lines illustrate the spacial distance between the true- and the reference
    paths along the sight-lines, respectively. \textit{Bottom}: GW strain as a function of time. The {\it turquoise} curve shows the GW signal
    for the observed BBH inspiral (true path), where the {\it orange} shows what the isolated BBH merger signal would look like (reference path).
    The two signals are shifted by the light crossing time between the true and the reference BBH paths, i.e. the time it takes the GWs to travel
    along the green lines. This gives rise to a unique observable GW phase shift that can be directly mapped to the BBH formation and environment.}
    \label{fig:GWPH_ill_1}
\end{figure*}


In this study we perform post-Newtonian ($\PN$) N-body simulations to show that BBHs formed through binary-single
interactions occurring in dense stellar clusters naturally lead to a well-defined population of GW sources with a potentially measurable GW
phase shift caused by the presence of the third object. Binary-single interactions are common occurrences in stellar clusters and consist of an initially wide BBH that interacts gravitationally with a nearby single BH \citep[e.g.][]{2018PhRvD..97j3014S}, leading to a 3-body interaction that may result in a BBH merger. This is illustrated in Fig.~\ref{fig:GWPH_ill_1}, which shows an
illustration of different dynamical interactions in a cluster environment ({\it upper left}), a BH binary-single interaction
producing a GW-driven merger ({\it upper right}), and the corresponding GW phase-shifted signal
caused by the presence of the third-object ({\it lower panel}). Several dynamical pathways
can lead to BBH mergers in clusters \citep[e.g.][]{2020PhRvD.101l3010S}, but the pathway relevant for clearly producing phase-shifted
sources observable by ground-based detectors is characterized by two of the three interacting BHs inspiralling and merge while the third is
still bound (Fig. \ref{fig:GWPH_ill_1} upper right and Fig. \ref{fig:dist_dphifp}),
also referred to as a {\it 3-body merger} (see Methods).
Such 3-body mergers constitute $\sim 10\%$ of the BBH mergers from globular clusters (GCs, which are considered to be gas-poor) \citep{2018PhRvD..97j3014S} and
dominate the fraction of eccentric sources, which implies that GW phase-shifted sources will form in notable
numbers with observable prospects for both current ground-based GW detectors (LIGO/Virgo/KAGRA) \citep{LIGOScientific:2014pky,VIRGO:2014yos,KAGRA:2018plz,2018LRR....21....3A} and future (Einstein Telescope (ET) \citep{2020JCAP...03..050M}, Cosmic Explorer (CE) \citep{2023arXiv230613745E}), as well as space-borne detectors (DECIGO/TianQin/Taiji
\citep{2011CQGra..28i4011K, 2016CQGra..33c5010L, 10.1093/nsr/nwx116, 2020PhRvD.101j3027L} (deci-Hertz)
and LISA \citep{2017arXiv170200786A} (milli-Hertz)).

In Sec \ref{AP_Analytical Framework} we describe the analytical framework and numerical setup employed in this work. We show and discuss our results in Sec. \ref{sec:Results} and summarise in Sec. \ref{sec:Conclusion}. 

\section{Methods}

\subsection{Analytical Framework}\label{AP_Analytical Framework}

The GW phase shift in 3-body mergers arises from the BBH being bound to the single BH while merging, and therefore moving on a
curved orbit with a time-varying velocity along the line-of-sight (LOS) relative to the observer.
This creates a time-dependent Doppler shift of the GW waveform, which translates to a GW phase shift that can be mapped to the
dynamics of the 3-body system \citep{2024arXiv240305625S}, analogous to binary pulsar
timings \citep[e.g.][]{2017ApJ...834..200M, 2024arXiv240305625S}.

In the limit where the acceleration ${\bf a}$ of the BBH center-of-mass (COM) near merger from the presence of the third body can be
assumed constant, the maximum displacement in time between a GW signal sent from the accelerated BBH path and a non-accelerated path as
seen by a distant non-cosmological observer is given by \citep[e.g.][]{2024arXiv240804603H},
\begin{equation}
\tau(t) =\frac{1}{2c} |{\bf a}| t^2 = \frac{1}{2} \frac{Gm}{c}\frac{t^2}{R^2}.
\label{eq:tau}
\end{equation}
Here we assume all three BHs have equal mass $m$, $c$ denotes the speed-of-light, and $R$ refers to the distance between the
BBH COM and the single BH at the time of merger.
This $\tau$ corresponds to the time it takes a light pulse to travel along one
of the green lines shown in Fig. \ref{fig:GWPH_ill_1}; a time that often is referred to as the {\it Rømer Delay}. The time-dependent GW phase shift can be approximated by the light crossing time between the perturbed (accelerated path, turquoise in Fig. \ref{fig:GWPH_ill_1}) and the
un-perturbed (linear path, orange in Fig. \ref{fig:GWPH_ill_1}), defined by $\tau$, divided by the orbital time of the inspiralling BBH, $T$, times $2\pi$ (see Fig. \ref{fig:GWPH_ill_1}) \citep{2024arXiv240305625S}:
\begin{align}
	\Delta{\phi}(t) & \approx 2\pi {\tau(t)}/{T(t)}.
\end{align}
Explicitly, using
\begin{equation}
T(t) = 2\pi\sqrt{a(t)^3/2Gm},
\end{equation}
the phase shift becomes:

\begin{align}
	\Delta{\phi}(t) & \approx 2\pi \frac{{\tau(t)}}{T(t)} \approx \frac{\sqrt{2}}{2}\frac{G^{3/2}}{c} \frac{m^{3/2}}{R^2} \times \frac{t^2}{a(t)^{3/2}} \,,
    \label{eq:dphi_general}
\end{align}
where $T(t)$ and $a(t)$ denote the BBH orbital time and the BBH semi-major axis (SMA), respectively, at time $t$.

Using the analytical framework of \cite{Peters64}, hereafter referred to as {\it Peters64}, for relating the time $t$ to the BBH's
eccentricity, $e$, and SMA, $a$, as well as the relation for $a(e)$, one can also write the phase shift as a function of the BBH's eccentricity.
For this we approximate the
merger time by the expression given by Peters64,
\begin{equation}
t_e \approx \frac{3}{85}\frac{c^5}{G^{3}}\frac{a^{4}  (1-e^{2})^{7/2} }{2m^3},
\label{eq:tm_ecc}
\end{equation}
where $e$ is the eccentricity of the BBH. By now using the relation between
$a$ and $e$ as presented in Peters64,
\begin{equation}
a(e) = \frac{C_0e^{12/19}}{(1-e^2)} \times g(e),
\label{eq:AP_ae}
\end{equation}
where
\begin{equation}
g(e) = \left(1+121e^2/304\right)^{870/2299},
\end{equation}
and $C_0$ is a constant that depends on
the initial conditions, $a_0, e_0$, we can now replace the dependence of $a$ with eccentricity $e$ in the above relations.
For determining $C_0$ we note that the 3-body BBH mergers we consider all start with high eccentricity ($e_0 \approx 1$),
from which it follows that the constant $C_0 \approx 2r_{0}/g(1)$, where $r_0  = a_0(1-e_0)$.
With this constant we can now write the relation $a(e)$ as, 
\begin{equation}
a(e) \approx \frac{2r_0e^{12/19}}{(1-e^2)}\frac{g(e)}{g(1)},\ (e_0 \approx 1).
\label{eq:ae_e1lim}
\end{equation}
By substituting this relation between $a$ and $e$ into Eq. \ref{eq:dphi_general} and Eq. \ref{eq:tm_ecc}, one finds
\begin{align}
    \Delta{\phi}(e) & \approx \frac{144}{85^{2}g(1)^{13/2}} \frac{c^{9}}{G^{9/2}} \times \frac{1}{R^2}\frac{r_0^{13/2}}{2^{3/2}m^{9/2}} \nonumber\\
	 	    &  \times e^{78/19}(1-e^2)^{1/2}g(e)^{13/2},
    \label{eq:AP_Dphi_e}
\end{align}
Here, $r_0$ is the peri-center distance of the merging BBH at the time of formation
during the interaction. This $r_0$ maps to a corresponding GW peak frequency as \citep[e.g.][]{2018PhRvD..97j3014S}
\begin{equation}
f_0 \approx \frac{1}{\pi}\sqrt{{2Gm}/{r_0^{3}}},
\label{eq:fp_rp}
\end{equation}
which can be used as a proxy for judging if the BBH is likely to be in the observable band or not (see also \cite{2024ApJ...969..132V}). It should be noted that the definition of the eccentricity in this work is based on the peak harmonic of the orbital frequency \cite{2003ApJ...598..419W, 2024ApJ...969..132V}. A convenient aspect of writing Eq. \ref{eq:AP_Dphi_e} in terms of $r_0$ is that $r_0$ only varies very little during the eccentric inspiral
phase, in contrast to $a,e$. This implies that $r_0$ can be defined relatively accurately without specifying exactly at what point in
time it is defined, as long as the BBH is still eccentric enough, or equivalently close enough to its initial assembly.
We elaborate further on this in the Sec. \ref{sec:Numerical Methods}.

\begin{figure*}
    \centering
    \includegraphics[width=.99\textwidth]{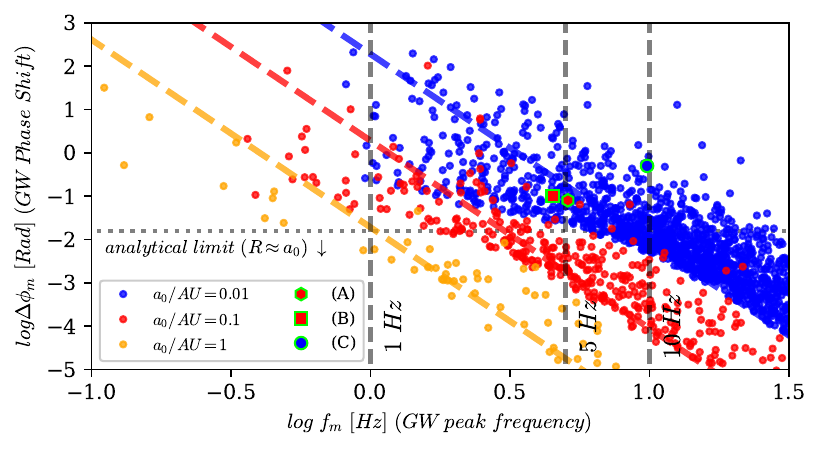}
    \caption{{\bf Distribution of maximum GW phase shifts from 3-body BBH mergers.} Results from $\PN$ simulations between a
    BBH and a single incoming BH that concludes with two of the three BHs merging while the third BH is still bound
    (see Fig. \ref{fig:GWPH_ill_1} and Fig. \ref{fig:orbit_ex}).
    Each dot shows the {\it maximum observable GW phase shift} (y-axis) for each of these mergers, as a function of the corresponding GW
    peak frequency (bottom x-axis) or merger time (top x-axis), derived using the methods outlined in
    Sec. \ref{AP_Analytical Framework}. Each colour refers to a different
    SMA of the initial BBH before interaction: $\sim 1$ AU ({\it orange}), $\sim 0.1$ AU ({\it red}), and $\sim 0.01$ AU ({\it blue}).
    The {\it dashed coloured lines} illustrate the asymptotic limit $\Delta{\phi} \propto f^{-13/3}$,
    where the {\it horizontal dotted line} indicates the analytical maximum value for $\Delta{\phi}$ assuming the distance between the BBH and the
    single BH, $R$, is similar to the initial SMA, $a$ (see Methods).    
    The three highlighted examples, {\bf (A,B,C)}, are shown and studied further in Fig. \ref{fig:orbit_ex}.
    }
    \label{fig:dist_dphifp}
\end{figure*}

Considering Eq. \ref{eq:AP_Dphi_e}, we see that the only part of the function that relates to the time-evolving BBH is the function
\begin{equation}
F(e) = e^{78/19}(1-e^2)^{1/2}g(e)^{13/2},
\end{equation}
which has a maximum that defines the maximum of the time-evolving GW phase shift. The eccentricity that leads to this
maximum, referred to as $e_m$, is $e_m \approx 0.95$ (see also \citep{2024arXiv240305625S})
with a corresponding GW peak frequency, $f_{m} = f(e_m)$, that
follows from combining Eq. \ref{eq:ae_e1lim} and \ref{eq:fp_rp},
\begin{equation}
f \approx {f_0} \times \left(\frac{2e^{12/19}}{1+e}\frac{g(e)}{g(1)} \right)^{-3/2}.
\label{eq:fmf0_e}
\end{equation}
The value for $f_{m}$ can be below the observable band (see Fig. \ref{fig:dist_dphifp}), and to quantify
the maximum measurable value for $\Delta{\phi}$ when the BBH enters the band one therefore needs to
rewrite Eq. \ref{eq:AP_Dphi_e} in terms of $f$ instead of $e$. However, it is not algebraically possible to
express $e$ as a function of $f$ using Peters64's relations.
For this, we therefore need to make an approximation in the limit $e \approx 1$ limit for which we can write
Eq. \ref{eq:fmf0_e} as $(f/f_0)^{-2/3} \approx e^{12/19}$ that leads to
$e \approx (f_0/f)^{19/18} \approx (f_0/f)$. Substituting this into Eq. \ref{eq:AP_Dphi_e} one finds,
\begin{align}
\Delta{\phi}(f) & \approx \frac{c^{9}G^{-7/3}}{2\pi^{13/3}} \left(\frac{5}{256}\right)^{2} \times \frac{1}{R^2}\frac{2^{2/3}}{m^{7/3}}\nonumber\\
                & \times f^{-13/3} \times (1+({f_0}/{f}))^7(1-({f_0}/{f}))^{1/2}.
\label{eq:AP_Dphi_f}
\end{align}
The $\propto f^{-13/3}$ factor correctly defines the asymptotic limit ($f \gg f_0$), as this is the solution to
the well studied circular case \citep{2017ApJ...834..200M}. Despite the approximation used between $e$ and $f$ above,
we find this expression to correctly capture the GW phase shift evolution within a factor of unity of our considered
eccentric inspiraling BBH, as a function of $f$ (see also \citep{2024arXiv240305625S}).
To summarise, given $r_0$ and $R$ for each individual scattering,
one can now estimate the maximum GW phase shift $\Delta{\phi}(e_m)$ using Eq.~\eqref{eq:AP_Dphi_e} and the corresponding $f_m$
using Eq.~\eqref{eq:fmf0_e}.

The analytical framework of Eqs. \ref{eq:AP_Dphi_e} and \ref{eq:AP_Dphi_f} builds on the 2.5 $\PN$ orbit-averaged
equations presented by Peters64. Despite these only including the lowest order $\PN$ terms and using Newtonian definitions of the orbital elements \citep[e.g.][]{2020MNRAS.495.2321Z}, these equations still offer very useful insight into the general scalings of the problem, as well as a reliable and fast way of estimating the GW phase shift of 3-body mergers. The phase shift that we consider and is generally considered in the literature \citep[e.g.][]{2017ApJ...834..200M, Vijaykumar2023-qg, 2018PhRvD..98f4012R} simply originates from the acceleration of the binary COM, as is evident from Eq. \ref{eq:tau}. Therefore, our scalings with masses and distance are fully consistent with this framework. Through the Peters64 equations, effects from the evolving eccentricity are only included in our model at the lowest order. While the Peters64 inspiral time is known to differ by up to a factor of a few from the exact general relativistic value, particularly at high eccentricities \citep{2025PhRvD.112b4012F}, we note that the distributions and results presented in Fig. \ref{fig:dist_dphifp} are fundamentally determined by the proximity of the third object to the binary, and are thus not sensitive to the precise inspiral time prescription. However, if desired, our framework can be followed up using more sophisticated analysis methods (see Sec. \ref{sec:Results} and \citep{2024arXiv240305625S}). 

For deriving these relations, we make use of the formalism from {\citep{2024arXiv240305625S, 2024arXiv240804603H}}, which presented analytical closed-form
solutions to the limit where the acceleration of the eccentric inspiraling BBH is assumed constant, also referred to here as the {\it linear limit}. In \cite{2024arXiv240804603H} we showed explicitly that the maximum phase shift and the overall morphology can be accurately estimated in the linear limit for the 3-body interactions considered here, even for large outer eccentricities.

Furthermore, we focus on the highest possible GW phase shift, denoted by $\Delta{\phi}$, which is related to the observer-dependent value given by $\delta{\phi} \approx \Delta{\phi} \times \cos(i)\sin(j)$, where $i$
is the angle in the plane of the `outer binary' (the binary composed of the BBH and the single BH) between the
LOS of the observer and the line connecting the location of BBH merger and the COM of the 3-body system, and
$j$ is the angle between the LOS and the angular momentum vector of the `outer binary' \citep[e.g.][]{2017ApJ...834..200M, 2024arXiv240305625S}. We make this decision in order to isolate the effects that are intrinsic to the phase shift.


\subsection{Dynamical Constraints}\label{Sec:Dynamical Constraints}

In our considered linear limit it appears from Eq. \ref{eq:AP_Dphi_e} that any GW phase shift can be achieved,
as long as $R$ and $r_0$ are chosen to be small and large enough, respectively. However, by imposing a dynamical
stability to the system, our linear limit also
implies a maximum value for $\Delta{\phi}$ as a function of $m$ and $R$. To see this we note that $r_0$ cannot take any value, but has
to be small enough that the radiated energy over the first peri-center passage,
\begin{equation}
\Delta{E}_{0} \approx \frac{85{\pi}G^{7/2}}{12\sqrt{2}c^{5}} \frac{m_1^2m_2^2\sqrt{m_{12}}}{r_0^{7/2}}.
\label{eq:DE_GW}
\end{equation}
leads to an orbital energy
\begin{equation}
E_{0} = \frac{Gm_1m_2}{2a_0},
\label{eq:AP_E0}
\end{equation}
that has a corresponding SMA $a_0$ that is smaller than the BBH Hill radius, $R_H$,
\begin{equation}
R_H \sim R((m_1+m_2)/m_3)^{1/3},
\label{eq:RH}
\end{equation}
by a factor $\beta$, such that $a_0 \lesssim \beta R_H$. Here, $m_1$ and $m_2$ are the binary component masses (with $m_{12} = m_1 + m_2$) and $m_3$ is the tertiary mass. By now equating $E_{0}$ and $\Delta{E}_{0}$, and impose the requirement
$a_0 = \beta R_H$ for stability, one finds the following maximum possible value for $r_0$, defined here as $r_H$,
\begin{equation}
r_H \approx \left(\frac{85\pi}{6\sqrt{2}} \frac{G^{5/2}}{c^{5}} \frac{\beta R m_1m_2m_{12}^{5/6}}{m_3^{1/3}}\right)^{2/7}.
\label{eq:rH_RH}
\end{equation}
The corresponding maximum GW phase shift can now be found be substituting $r_H$ from the above Eq. \ref{eq:rH_RH} into Eq. \ref{eq:AP_Dphi_e},
\begin{equation}
    \Delta{\phi}_H(e) \approx 10^{-2} \times \left(\frac{\beta}{0.1}\right)^{13/7} \left(\frac{m/M_{\odot}}{R/AU}\right)^{1/7} F(e),
    \label{eq:maxphi_rH}
\end{equation}
where we set $m=m_1=m_2=m_3$. As seen here, the maximum value estimated from the characteristic length- and mass-scales of the problem in this
linear limit is therefore relatively small, as well as only weakly dependent on the BH mass and initial orbital seperation.



\subsection{Numerical Methods}\label{sec:Numerical Methods}

The scatterings presented in this work were generated with a $\PN$ few-body code that includes 1- and 2-$\PN$ terms which lead to precession as well as the dissipative 2.5-$\PN$ term \citep{Blanchet06, Blanchet14}. For each of our scatterings we initiate a circular BBH with a given SMA $a$,
and sample the distribution of incoming single BHs at infinity according to the system we consider.
From the distribution at infinity, we then propagate each single close to the binary using Kepler's
equations following well known procedures \citep{Hut:1983js, Samsing14}.
At this closer distance we then initiate our $\PN$ few-body code, and let it run for a total of $1000$ initial BBH orbital times.
This resulted in a few percent of inconclusive scatterings, that mainly originates from scatterings for
which the third object is sent out on an almost unbound orbit \citep[e.g.][]{1983AJ.....88.1549H, 2018MNRAS.476.1548S}.
For each BBH that merges during the interaction, i.e. for each 3-body merger, we measure the initial peri-center distance
$r_0$ and the distance from the BBH COM to the third object $R$ at merger, from which we theoretically estimate the GW phase shift using the
formalism from the above Sec. \ref{AP_Analytical Framework}.
The distance $r_0$ is measured numerically when the inspiraling BBH is subject to a tidal force
from the third object that is $< 10^{-3}$ compared to its own binding force. In this way we make sure that tides no longer
impact the BBH evolution. The distance $r_0$ stays approximately constant until the merging BBH starts circularizing,
which implies $r_0$ can be measured relatively unambiguously.

The orange coloured paths seen in Fig. \ref{fig:GWPH_ill_1} (and further ahead in Fig. \ref{fig:orbit_ex}), which illustrate the trajectories
the BBHs would take if the third object was not there, are derived by evolving the $\PN$ equations-of-motion backwards from merger without
the third object. This novel technique, as further described in \cite{2024arXiv240305625S}, allows for an accurate way of numerically estimating observable Doppler and tidal effects in the GW signal induced by the presence of the third object.

\section{Results \& Discussion}\label{sec:Results}

\begin{figure*}
    \centering
    \includegraphics[width=.99\textwidth]{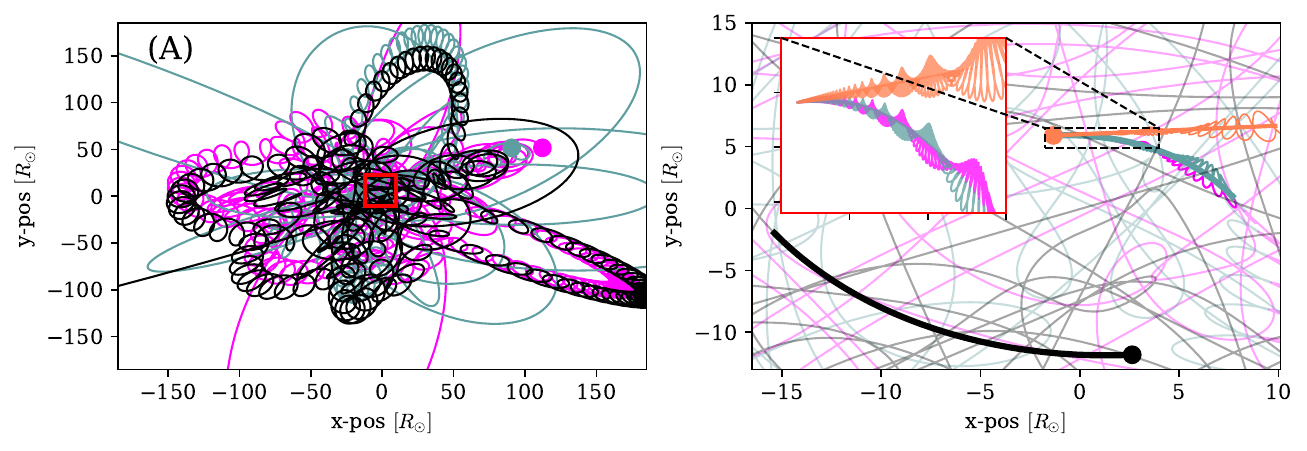}
    \includegraphics[width=.99\textwidth]{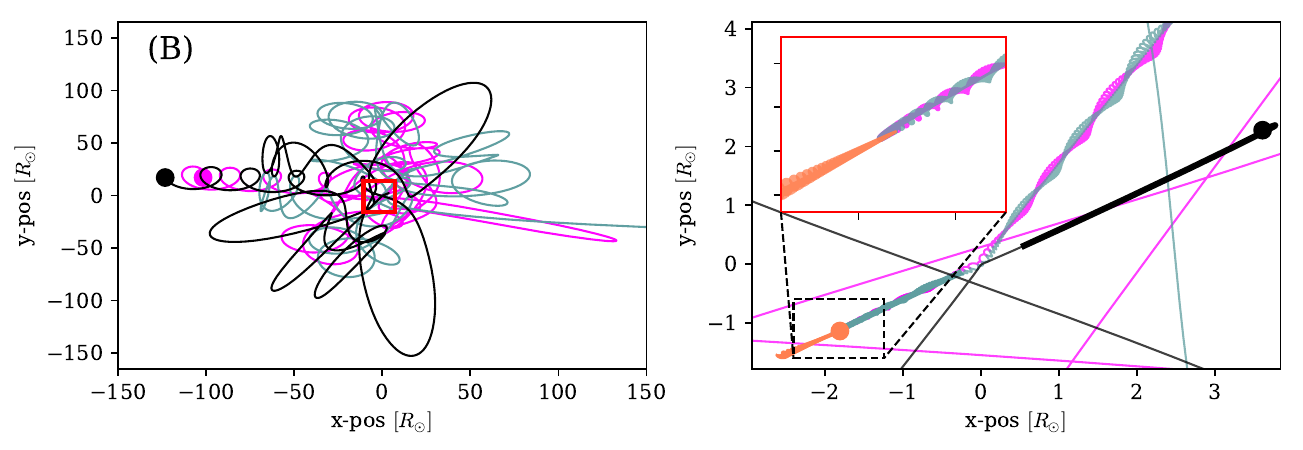}
    \includegraphics[width=.99\textwidth]{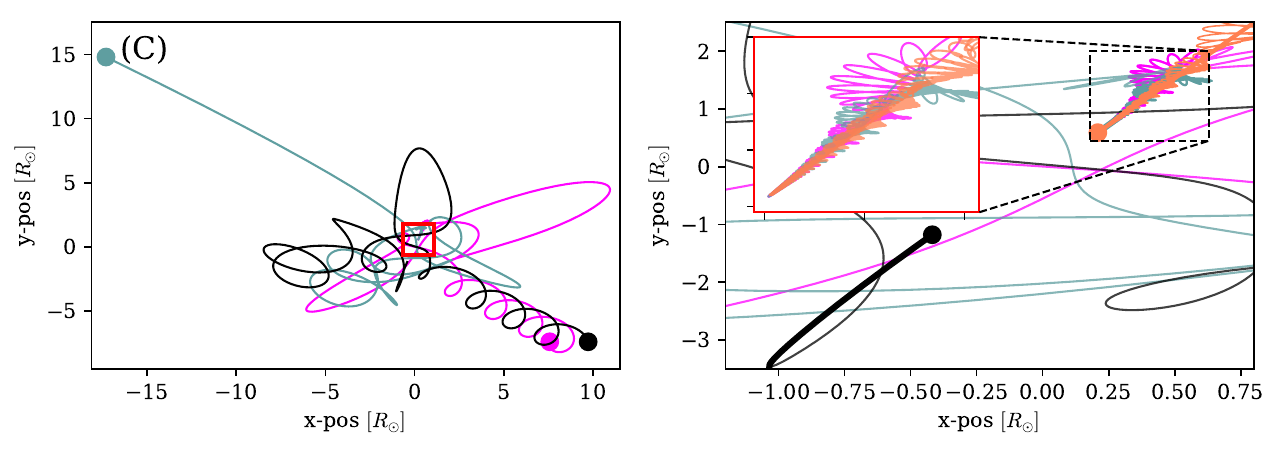}
    \caption{{\bf Examples of 3-body BBH mergers resulting in significant GW phase shifts.}
    The shown interactions involve BHs with equal mass $m = 20M_{\odot}$, and initial SMA
    $a = 0.1$ AU for case {\bf (A)} and {\bf (B)}, where case {\bf (C)} has $a = 0.01$ AU. The filled-circles in the left figures
    indicate the initial positions of the BHs, where the filled-circles in the right figures show the end positions at merger.
    The orange lines illustrate the trajectory the merging BBHs would take if the third BH was not there (see Fig. \ref{fig:GWPH_ill_1}).
    Common for the cases that give rise to significant GW phase shifts is that the chaotic nature of the 3-body problem
    brings the BBH close to the remaining bound single BH near merger. This is clearly seen in these examples.
    GW phase shifts and corresponding peak frequencies for (A,B,C) are shown in Fig. \ref{fig:dist_dphifp}.}
    \label{fig:orbit_ex}
\end{figure*}

\begin{figure*}
    \centering
    \includegraphics[width=.49\textwidth]{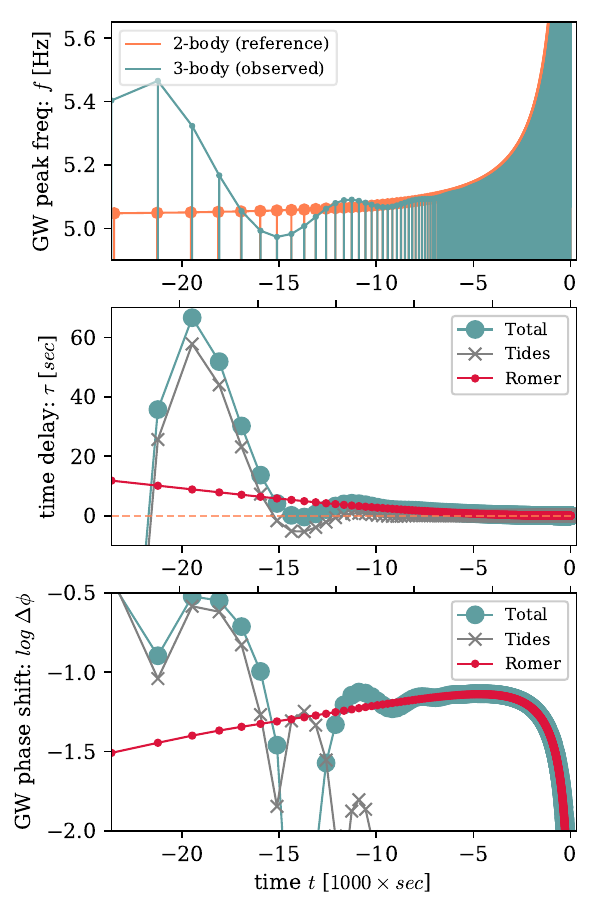}
    \includegraphics[width=.49\textwidth]{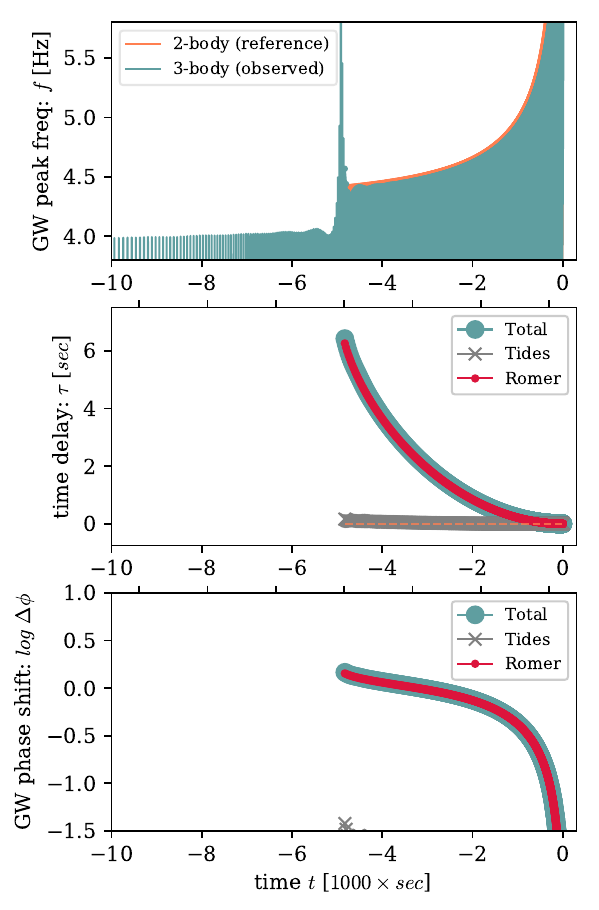}
    \caption{ {\bf GW Modulations in 3-body interactions.}
    Numerically derived differences, as seen by a distant non-cosmological observer,
    between the true inspiraling BBH and its reference BBH for the two examples; ${\bf (A)}$ ({\it left column})
    and ${\bf (B)}$ ({\it right column}) introduced in Sec. \ref{sec:Results}.
    For both columns the {\it top plot} shows the evolution of BBH peak frequency $f$ (the blue and orange
    lines show results for the perturbed 3-body BBH and its 2-body
    reference BBH, respectively), the {\it middle plot} the time delay $\tau$ (the blue line shows the total,
    the grey line shows the contribution from tides, where the red line is the contribution from the BBH COM acceleration),
    and in the {\it bottom plot} is shown the corresponding GW phase shift (same notation as for the middle plot).
    Our analytical framework from Sec. \ref{AP_Analytical Framework} describes the evolution shown here
    with red lines (COM acceleration), where the deviations away from this are coursed by semi-chaotic
    tidal interactions between the single BH and the BBH.
    }
    \label{fig:AP_acc_gen_fig}
\end{figure*}


We now consider results from controlled $\PN$ binary-single interactions between BBHs with initial
SMA ${a = 0.01, 0.1, 1~\mathrm{AU}}$, respectively, and an incoming BH, all with equal mass $m=20~M_{\odot}$.
These values are characteristic for BHs interacting in dense stellar systems, as these map to stellar clusters
with escape velocities, $v_e$, of $\sim 30\ \mathrm{km\,s}^{-1}$ (globular cluster),
$\sim 90\ \mathrm{km\,s}^{-1}$ (dense stellar cluster), $\sim 300\ \mathrm{km\,s}^{-1}$ (nuclear star cluster),
for $a = 1, 0.1, 0.01~\mathrm{AU}$, respectively \citep[e.g.][]{2016ApJ...831..187A}. For every 3-body merger that forms, we measure numerically $r_0$ and $R$, from which we estimate the maximum GW phase shift and corresponding GW peak frequency, as outlined in Sec. \ref{AP_Analytical Framework}. While the smallest sampled SMA may appear optimistic for purely dynamical three-body encounters in NSCs, interactions with the gas in AGN discs can make such compact configurations possible in that context (see \cite{2025MNRAS.539.1501R, 2026ApJ...998..244T}).

\begin{figure*}
    \centering
    \includegraphics[width=.99\textwidth]{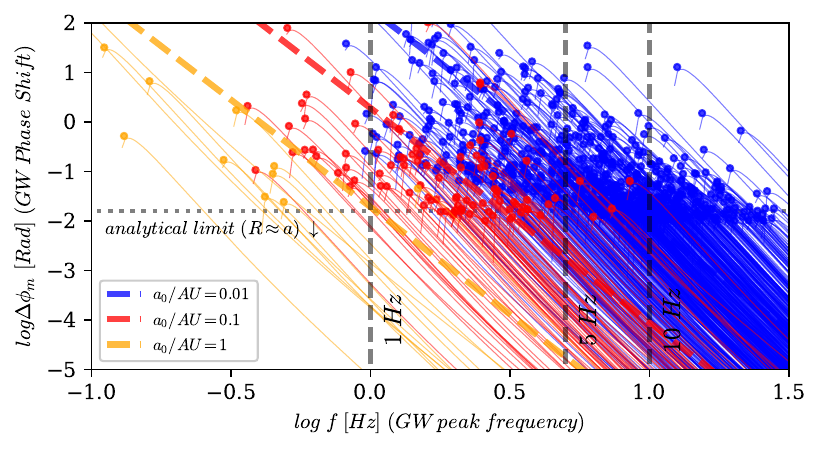}
    \caption{{\bf Evolution of GW phase shift with GW peak frequency.} The figure shows evolutionary tracks of the GW phase shift as a function
    of the GW peak frequency, $f$, from its initial value at formation, $f_0$, derived using Eq. \ref{eq:AP_Dphi_f}.
    The data is the same as the one shown in Fig. \ref{fig:dist_dphifp}, but only including the GW sources with a
    maximum GW phase shift of $>10^{-2}$. The filled-circles are identical to the ones shown in Fig. \ref{fig:dist_dphifp},
    and illustrate correctly (within our analytical approximations) the maximum value for the GW phase shift along each track.}
    \label{fig:evolv_dphif}
\end{figure*}

Fig. \ref{fig:dist_dphifp} shows our numerically derived distributions of $f_m$ and $\Delta{\phi}(e_m)$.
As seen, the chaotic nature of the binary-single problem leads to GW sources with notable GW phase
shifts, especially in the range between $1-10$ Hz, accessible to third-generation (3G) ground-based GW observatories such as ET and CE. For instance, $\sim 20\%$ of the population with $a_0 = 0.1$ AU (red) has a maximum phase shift above 0.01 radians. This result is non-trivial, as the analytical estimate for the maximum GW phase shift of Eq. \ref{eq:maxphi_rH}
predicts a near universal upper limit of $\Delta{\phi}_H(e) \sim 10^{-2} \times (m[M_{\odot}]/R[\mathrm{AU}])^{1/7}$
radians \citep{2024arXiv240305625S}.

To understand what kind of 3-body evolutions are able to bypass the analytical
limit of $\sim 10^{-2}$ radians, we consider the three distinct examples shown in Fig. \ref{fig:orbit_ex}. In {\bf (A)}, we show a highly chaotic and long-lived interaction, that concludes with an endstate charaterized
by the BBH passing the bound single near closest approach on a clearly curved orbit.
In {\bf (B)}, we show an example where the BBH merges right after turning around to directly move towards the bound single.
This creates a large offset between the reference- and the true trajectory. Note also how the single BH scatters off the BBH near
the 3-body COM $(0,0)$ as it inspirals. This creates large extra perturbations to the GW signal.
In {\bf (C)}, we show a case where the merging BBH is heading directly towards the bound single BH. Here there is no significant curvature,
but the acceleration along its direction of motion towards the single BH creates the GW phase shift.

As follows from Eq. \ref{eq:AP_Dphi_e}, the only important parameter that can freely vary during interactions, and also lead to a large GW phase
shift at a given $f$, is the distance $R$. In general, the GW sources we see forming with the largest GW phase shift are therefore all characterized by having the remaining single BH much closer at merger than expected from the initial binary separation. As illustrated, this all follows naturally from the chaotic nature of the 3-body problem, which further opens up for
new directions in performing, classifying and identifying GW phase-shifted sources in $\PN$ few-body scatterings.

While our analytical descriptions in Sec. \ref{AP_Analytical Framework} accurately describe the amplitude and overall morphology of the phase shift, more insight could be gained into the rich information encoded these phase-shifted sources by computing the dynamics of these systems numerically. Numerical methods can be used to resolve the fine and often non-linear
differences between the true observed BBH and its corresponding reference GW signal. As an example, Fig. \ref{fig:AP_acc_gen_fig} shows numerically derived quantities for
example ${\bf (A)}$ and ${\bf (B)}$ presented in Fig. \ref{fig:orbit_ex}.
The quantities are derived by comparing the GW signal a distant non-cosmological observer would see from the
true BBH evolving near the third object with a reference BBH created by backwards simulating the $\PN$ equations without the third
object (Sec. \ref{sec:Numerical Methods} and \cite{2024arXiv240305625S}).
In case of example ${\bf (A)}$, one clearly sees periodic modulations in both the GW peak frequency ({\it top}), time delay ({\it middle}),
and GW phase shift ({\it bottom}) below $\sim 5000$ seconds prior to merger. These originate from the time-dependent tidal coupling
between the single BH and the precessing BBH as its inspirals. When
the BBH becomes compact enough the tidal coupling fades away, after which the GW phase shift becomes dominated by
the BBH COM acceleration. As seen, this part in the GW phase shift until merger, is very smooth and follows
accurately our formalism presented in Sec. \ref{AP_Analytical Framework}.
In example ${\bf (B)}$ is seen a clear strong perturbation of the GW signal $\sim 5000$ seconds prior to merger.
This is due to the fact that the single BH scatters off the BBH close to merger (see Fig. \ref{fig:orbit_ex}), which
leads to an almost impulsive change in the BBH orbital parameters. For 3G detectors operating near $5$ Hz, such highly non-linear
outcomes should be present, and would provide an enormous amount of unique information about the BBH formation
beyond the corresponding GW phase shift and linear predictions.

\begin{figure}
    \centering
    \includegraphics[width=0.5\textwidth]{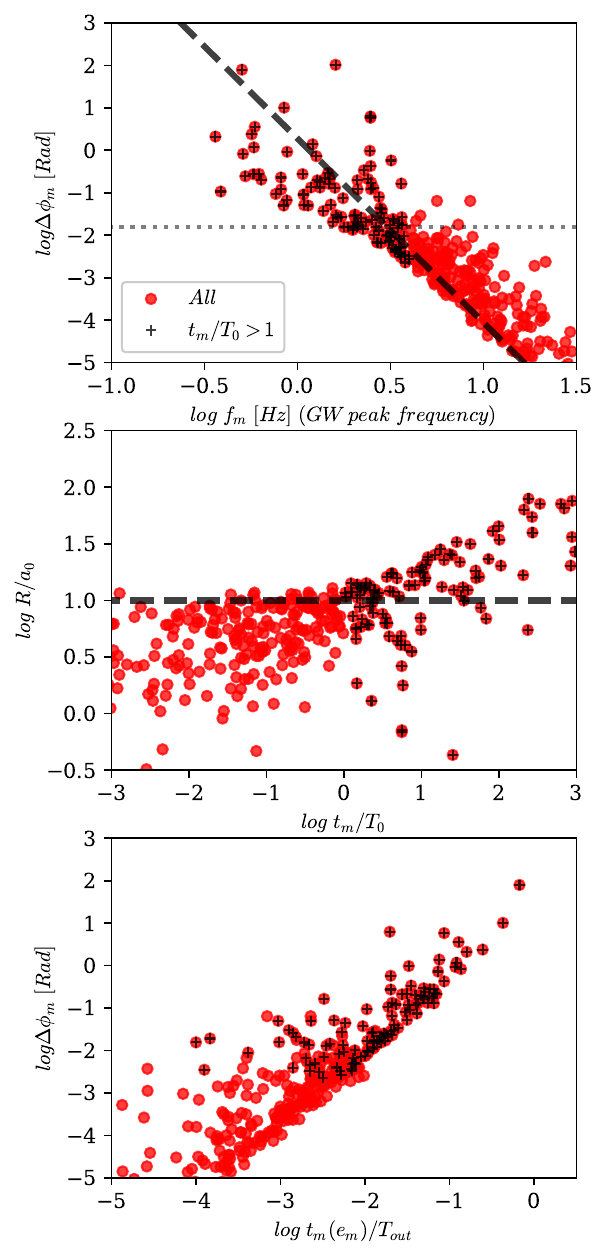}
    \caption{\textbf{Characteristics of phase-shifted 3-body BBH mergers.} \textit{Top:} the GW phase shift as a function of the peak GW frequency (see Fig. \ref{fig:dist_dphifp}) for the population with $a_0=0.1 AU$. The $+$-marked systems are those binaries whose time to merger at assembly ($t_m$) exceeds the orbital period of the initial binary ($T_0$). The black dashed line is the asymptotic limit $f^{-13/3}$ and the dotted line is the expected analytical upper threshold according to our analytical model. \textit{Middle:} the distance $R$ at merger between the binary and third body, relative to the semi-major axis of the initial binary, $a_0$. \textit{Bottom:} the maximum GW phase shift as a function of the merger time at the maximum (where $e=e_m$), relative to the outer orbital period of the tertiary in the 3-body merger ($T_\mathrm{out}$).}
    \label{fig:subpopulation_0.1AU}
\end{figure}

Fig. \ref{fig:evolv_dphif} shows the $\Delta{\phi}(f)$ evolution of the inspiraling BBHs from our data
presented in Fig. \ref{fig:dist_dphifp}, with a maximum $\Delta{\phi} > 10^{-2}$ radians
evolved using Eq. \ref{eq:AP_Dphi_f} from their initial formation frequency $f_0$, towards higher $f$. The dots at the
individual peaks are identical to $\Delta{\phi}(e_m)$, $f_m$, and derived
from Eq. \ref{eq:AP_Dphi_e} and Eq. \ref{eq:fmf0_e} (the slight offsets are due the approximations we here made for $\Delta{\phi}(f)$).
As seen, several of the BBH sources that peak at low $f$ are still able to enter higher frequency bands but with a smaller $\Delta{\phi}$.
Irrespective of 3G GW detectors (ET, CE) will operate down to $1$ or $5$ Hz, this clearly shows that GW sources
will enter, or even form in 3G GW detectors, and possibly also in current LVK-detectors with a significant GW phase shift.

Fig. \ref{fig:subpopulation_0.1AU} provides additional insight into the properties of the populations shown in Fig. \ref{fig:dist_dphifp}. We highlight on the population with an initial semi-major axis of 0.1 AU (red in Fig. \ref{fig:dist_dphifp}). The top panel highlights systems for which the merger time at \textit{assembly}, $t_m$, exceeds the orbital period, $T_0$, or the \textit{initial binary} in the scattering. This selection reveals two sub-populations. Systems with $t_m/T_0 > 1$ show phase shifts exceeding the analytical limit, but remaining below the asymptotic limit $f^{-13/3}$. More insight into this can be seen in the middle panel where we plot $t_m/T_0$ against the distance $R$ relative to semi-major axis of the initial binary $a_0$. This showcases that these systems correspond to larger separations $R/a_0$, as the third body can travel farther from the binary, leading to reduced phase shifts relative to the power law. Lastly, in the bottom panel we show the maximum phase shift as a function of the merger time $t_m$ at said maximum, occurring at eccentricity $e_m$, relative to the outer orbital period $T_\mathrm{out}$ of the eventual third body. This highlights that the merger time from the peak is typically much shorter than $T_\mathrm{out}$, justifying the assumption of constant acceleration.

\section{Conclusion}
\label{sec:Conclusion}
The study of GW phase shifts opens up tremendous possibilities for probing the origin and assembly mechanisms of individual GW sources. This work has shown that Rømer phase shifts with large amplitudes in accessible frequency ranges are a natural outcome of BBHs that merge through few-body interactions in stellar clusters. Given this and the sensitivity range of next-generation detectors, this establishes them as a serious candidate for detection in future GW signals if these dynamical environments contribute to the merger rate. Additionally, the rich dynamical complexity of these systems revealed in this work makes them a compelling target for further study.

Our findings may be used as a stepping stone to a more thorough understanding of phase-shifted binaries. Natural follow-up questions include characterising the population of systems that produce phase shifts above the detection threshold, and what their properties reveal about formation history. Additionally, dedicated injection studies with next-generation detectors across a range of GW frequencies would be a logical next step toward quantifying detectability of the phase shift and the orbital properties of these three-body systems.




\begin{acknowledgments}
The authors are grateful to Lorenzo Speri, Fabio Antonini and Daniel J. D'Orazio for useful discussions.
K.H, L.Z., P.S., J.T, and J.S. are supported by the Villum Fonden grant No. 29466, and by the ERC Starting
Grant no. 101043143 -- BlackHoleMergs led by J. Samsing.
M.Z. gratefully acknowledges funding from the Brinson Foundation in support of astrophysics research at the Adler Planetarium.
A.A.T. acknowledges support from the Horizon Europe research and innovation programs under the Marie Skłodowska-Curie grant
agreement no. 101103134.
\end{acknowledgments}

\bibliographystyle{aasjournal}
\bibliography{apssamp}

@ARTICLE{2025MNRAS.539.1501R,
       author = {{Rowan}, Connar and {Whitehead}, Henry and {Fabj}, Gaia and {Saini}, Pankaj and {Kocsis}, Bence and {Pessah}, Martin and {Samsing}, Johan},
        title = "{Prompt gravitational-wave mergers aided by gas in active galactic nuclei: the hydrodynamics of binary-single black hole scatterings}",
      journal = {\mnras},
         year = 2025,
        month = may,
       volume = {539},
       number = {2},
        pages = {1501-1515},
          doi = {10.1093/mnras/staf547},
archivePrefix = {arXiv},
       eprint = {2501.09017},
 primaryClass = {astro-ph.GA},
       adsurl = {https://ui.adsabs.harvard.edu/abs/2025MNRAS.539.1501R}
}

@ARTICLE{2026ApJ...998..244T,
       author = {{Tagawa}, Hiromichi and {Rowan}, Connar and {Tak{\'a}tsy}, J{\'a}nos and {Zwick}, Lorenz and {Hendriks}, Kai and {Han}, Wen-Biao and {Samsing}, Johan},
        title = "{Gravitational Wave Phase Shifts of Black Hole Mergers in AGN Disks}",
      journal = {\apj},
         year = 2026,
        month = feb,
       volume = {998},
       number = {2},
          eid = {244},
        pages = {244},
          doi = {10.3847/1538-4357/ae3a87},
archivePrefix = {arXiv},
       eprint = {2511.15193},
 primaryClass = {astro-ph.HE},
       adsurl = {https://ui.adsabs.harvard.edu/abs/2026ApJ...998..244T}
}

@ARTICLE{2021PhRvD.104j3011Y,
       author = {{Yu}, Hang and {Wang}, Yijun and {Seymour}, Brian and {Chen}, Yanbei},
        title = "{Detecting gravitational lensing in hierarchical triples in galactic nuclei with space-borne gravitational-wave observatories}",
      journal = {\prd},
         year = 2021,
        month = nov,
       volume = {104},
       number = {10},
          eid = {103011},
        pages = {103011},
          doi = {10.1103/PhysRevD.104.103011},
archivePrefix = {arXiv},
       eprint = {2107.14318},
 primaryClass = {gr-qc},
       adsurl = {https://ui.adsabs.harvard.edu/abs/2021PhRvD.104j3011Y}
}

@ARTICLE{2021PhRvL.126b1101Y,
       author = {{Yu}, Hang and {Chen}, Yanbei},
        title = "{Direct Determination of Supermassive Black Hole Properties with Gravitational-Wave Radiation from Surrounding Stellar-Mass Black Hole Binaries}",
      journal = {\prl},
         year = 2021,
        month = jan,
       volume = {126},
       number = {2},
          eid = {021101},
        pages = {021101},
          doi = {10.1103/PhysRevLett.126.021101},
archivePrefix = {arXiv},
       eprint = {2009.02579},
 primaryClass = {gr-qc},
       adsurl = {https://ui.adsabs.harvard.edu/abs/2021PhRvL.126b1101Y}
}

@ARTICLE{2025PhRvD.112b4012F,
       author = {{Fumagalli}, Giulia and {Loutrel}, Nicholas and {Gerosa}, Davide and {Boschini}, Matteo},
        title = "{Nonadiabatic dynamics of eccentric black-hole binaries in post-Newtonian theory}",
      journal = {\prd},
         year = 2025,
        month = jul,
       volume = {112},
       number = {2},
          eid = {024012},
        pages = {024012},
          doi = {10.1103/znmj-6wvt},
archivePrefix = {arXiv},
       eprint = {2502.06952},
 primaryClass = {gr-qc},
       adsurl = {https://ui.adsabs.harvard.edu/abs/2025PhRvD.112b4012F}
}

@ARTICLE{2003ApJ...598..419W,
       author = {{Wen}, Linqing},
        title = "{On the Eccentricity Distribution of Coalescing Black Hole Binaries Driven by the Kozai Mechanism in Globular Clusters}",
      journal = {\apj},
         year = 2003,
        month = nov,
       volume = {598},
       number = {1},
        pages = {419-430},
          doi = {10.1086/378794},
archivePrefix = {arXiv},
       eprint = {astro-ph/0211492},
 primaryClass = {astro-ph},
       adsurl = {https://ui.adsabs.harvard.edu/abs/2003ApJ...598..419W}
}

@article{KAGRA:2018plz,
    author = "Akutsu, T. and others",
    collaboration = "KAGRA",
    title = "{KAGRA: 2.5 Generation Interferometric Gravitational Wave Detector}",
    eprint = "1811.08079",
    archivePrefix = "arXiv",
    primaryClass = "gr-qc",
    reportNumber = "JGW-P1809243",
    doi = "10.1038/s41550-018-0658-y",
    journal = "Nature Astron.",
    volume = "3",
    number = "1",
    pages = "35--40",
    year = "2019"
}

@article{VIRGO:2014yos,
    author = "Acernese, F. and others",
    collaboration = "VIRGO",
    title = "{Advanced Virgo: a second-generation interferometric gravitational wave detector}",
    eprint = "1408.3978",
    archivePrefix = "arXiv",
    primaryClass = "gr-qc",
    doi = "10.1088/0264-9381/32/2/024001",
    journal = "Class. Quant. Grav.",
    volume = "32",
    number = "2",
    pages = "024001",
    year = "2015"
}

@article{LIGOScientific:2014pky,
    author = "Aasi, J. and others",
    collaboration = "LIGO Scientific",
    title = "{Advanced LIGO}",
    eprint = "1411.4547",
    archivePrefix = "arXiv",
    primaryClass = "gr-qc",
    doi = "10.1088/0264-9381/32/7/074001",
    journal = "Class. Quant. Grav.",
    volume = "32",
    pages = "074001",
    year = "2015"
}

@ARTICLE{2014barausse,
       author = {{Barausse}, Enrico and {Cardoso}, Vitor and {Pani}, Paolo},
        title = "{Can environmental effects spoil precision gravitational-wave astrophysics?}",
      journal = {\prd},
         year = 2014,
        month = may,
       volume = {89},
       number = {10},
          eid = {104059},
        pages = {104059},
          doi = {10.1103/PhysRevD.89.104059},
archivePrefix = {arXiv},
       eprint = {1404.7149},
 primaryClass = {gr-qc},
       adsurl = {https://ui.adsabs.harvard.edu/abs/2014PhRvD..89j4059B}
}

@ARTICLE{Fabj24,
Author = {Gaia Fabj and Johan Samsing},
Title = {Eccentric Mergers in AGN Discs: Influence of the Supermassive Black-Hole on Three-body Interactions},
Year = {2024},
journal = {arXiv e-prints},
eid = {arXiv:2402.16948},
pages = {arXiv:2402.16948},
url = {https://arxiv.org/pdf/2402.16948},
}

@ARTICLE{Vijaykumar2023-qg,
  title     = "Waltzing binaries: Probing the line-of-sight acceleration of
               merging compact objects with gravitational waves",
  author    = "Vijaykumar, Aditya and Tiwari, Avinash and Kapadia, Shasvath J
               and Arun, K G and Ajith, Parameswaran",
  journal   = "Astrophys. J.",
  publisher = "American Astronomical Society",
  volume    =  954,
  number    =  1,
  pages     = "105",
  month     =  sep,
  year      =  2023,
  copyright = "http://creativecommons.org/licenses/by/4.0/"
}

@ARTICLE{2019ApJ...885..135T,
       author = {{Trani}, Alessandro A. and {Spera}, Mario and {Leigh}, Nathan W.~C. and {Fujii}, Michiko S.},
        title = "{The Keplerian Three-body Encounter. II. Comparisons with Isolated Encounters and Impact on Gravitational Wave Merger Timescales}",
      journal = {\apj},
         year = 2019,
        month = nov,
       volume = {885},
       number = {2},
          eid = {135},
        pages = {135},
          doi = {10.3847/1538-4357/ab480a},
archivePrefix = {arXiv},
       eprint = {1904.07879},
 primaryClass = {astro-ph.GA},
       adsurl = {https://ui.adsabs.harvard.edu/abs/2019ApJ...885..135T}
}

@ARTICLE{2023MNRAS.523.4227A,
       author = {{Atallah}, Dany and {Trani}, Alessandro A. and {Kremer}, Kyle and {Weatherford}, Newlin C. and {Fragione}, Giacomo and {Spera}, Mario and {Rasio}, Frederic A.},
        title = "{Growing black holes through successive mergers in galactic nuclei - I. Methods and first results}",
      journal = {\mnras},
         year = 2023,
        month = aug,
       volume = {523},
       number = {3},
        pages = {4227-4250},
          doi = {10.1093/mnras/stad1634},
archivePrefix = {arXiv},
       eprint = {2211.09670},
 primaryClass = {astro-ph.GA},
       adsurl = {https://ui.adsabs.harvard.edu/abs/2023MNRAS.523.4227A}
}

@ARTICLE{2023MNRAS.524..426I,
       author = {{Iorio}, Giuliano and {Mapelli}, Michela and {Costa}, Guglielmo and {Spera}, Mario and {Escobar}, Gast{\'o}n J. and {Sgalletta}, Cecilia and {Trani}, Alessandro A. and {Korb}, Erika and {Santoliquido}, Filippo and {Dall'Amico}, Marco and {Gaspari}, Nicola and {Bressan}, Alessandro},
        title = "{Compact object mergers: exploring uncertainties from stellar and binary evolution with SEVN}",
      journal = {\mnras},
         year = 2023,
        month = sep,
       volume = {524},
       number = {1},
        pages = {426-470},
          doi = {10.1093/mnras/stad1630},
archivePrefix = {arXiv},
       eprint = {2211.11774},
 primaryClass = {astro-ph.HE},
       adsurl = {https://ui.adsabs.harvard.edu/abs/2023MNRAS.524..426I}
}

@ARTICLE{2022MNRAS.511.1362T,
       author = {{Trani}, Alessandro A. and {Rastello}, Sara and {Di Carlo}, Ugo N. and {Santoliquido}, Filippo and {Tanikawa}, Ataru and {Mapelli}, Michela},
        title = "{Compact object mergers in hierarchical triples from low-mass young star clusters}",
      journal = {\mnras},
         year = 2022,
        month = feb,
       volume = {511},
       number = {1},
        pages = {1362-1372},
          doi = {10.1093/mnras/stac122},
archivePrefix = {arXiv},
       eprint = {2111.06388},
 primaryClass = {astro-ph.HE},
       adsurl = {https://ui.adsabs.harvard.edu/abs/2022MNRAS.511.1362T}
}

@ARTICLE{2021MNRAS.504..910T,
       author = {{Trani}, A.~A. and {Tanikawa}, A. and {Fujii}, M.~S. and {Leigh}, N.~W.~C. and {Kumamoto}, J.},
        title = "{Spin misalignment of black hole binaries from young star clusters: implications for the origin of gravitational waves events}",
      journal = {\mnras},
         year = 2021,
        month = jun,
       volume = {504},
       number = {1},
        pages = {910-919},
          doi = {10.1093/mnras/stab967},
archivePrefix = {arXiv},
       eprint = {2102.01689},
 primaryClass = {astro-ph.HE},
       adsurl = {https://ui.adsabs.harvard.edu/abs/2021MNRAS.504..910T}
}

@ARTICLE{2023arXiv231213281T,
       author = {{Trani}, Alessandro Alberto and {Quaini}, Stefano and {Colpi}, Monica},
        title = "{Three-body encounters in black hole discs around a supermassive black hole: The disc velocity dispersion and the Keplerian tidal field determine the eccentricity and spin-orbit alignment of gravitational wave mergers}",
      journal = {arXiv e-prints},
         year = 2023,
        month = dec,
          eid = {arXiv:2312.13281},
        pages = {arXiv:2312.13281},
          doi = {10.48550/arXiv.2312.13281},
archivePrefix = {arXiv},
       eprint = {2312.13281},
 primaryClass = {astro-ph.HE},
       adsurl = {https://ui.adsabs.harvard.edu/abs/2023arXiv231213281T}
}

@ARTICLE{Samsing14,
       author = {{Samsing}, Johan and {MacLeod}, Morgan and {Ramirez-Ruiz}, Enrico},
        title = "{The Formation of Eccentric Compact Binary Inspirals and the Role of Gravitational Wave Emission in Binary-Single Stellar Encounters}",
      journal = {\apj},
         year = 2014,
        month = mar,
       volume = {784},
       number = {1},
          eid = {71},
        pages = {71},
          doi = {10.1088/0004-637X/784/1/71},
archivePrefix = {arXiv},
       eprint = {1308.2964},
 primaryClass = {astro-ph.HE},
       adsurl = {https://ui.adsabs.harvard.edu/abs/2014ApJ...784...71S}
}

@ARTICLE{Samsing2018,
       author = {{Samsing}, Johan and {MacLeod}, Morgan and {Ramirez-Ruiz}, Enrico},
        title = "{Dissipative Evolution of Unequal-mass Binary-single Interactions and Its Relevance to Gravitational-wave Detections}",
      journal = {\apj},
         year = 2018,
        month = feb,
       volume = {853},
       number = {2},
          eid = {140},
        pages = {140},
          doi = {10.3847/1538-4357/aaa715},
archivePrefix = {arXiv},
       eprint = {1706.03776},
 primaryClass = {astro-ph.HE},
       adsurl = {https://ui.adsabs.harvard.edu/abs/2018ApJ...853..140S}
}

@ARTICLE{Blanchet06,
       author = {{Blanchet}, Luc},
        title = "{Gravitational Radiation from Post-Newtonian Sources and Inspiralling Compact Binaries}",
      journal = {Living Reviews in Relativity},
         year = 2006,
        month = dec,
       volume = {9},
       number = {1},
          eid = {4},
        pages = {4},
          doi = {10.12942/lrr-2006-4},
       adsurl = {https://ui.adsabs.harvard.edu/abs/2006LRR.....9....4B}
}

@ARTICLE{Blanchet14,
       author = {{Blanchet}, Luc},
        title = "{Gravitational Radiation from Post-Newtonian Sources and Inspiralling Compact Binaries}",
      journal = {Living Reviews in Relativity},
         year = 2014,
        month = dec,
       volume = {17},
       number = {1},
          eid = {2},
        pages = {2},
          doi = {10.12942/lrr-2014-2},
archivePrefix = {arXiv},
       eprint = {1310.1528},
 primaryClass = {gr-qc},
       adsurl = {https://ui.adsabs.harvard.edu/abs/2014LRR....17....2B}
}

@ARTICLE{Peters64,
       author = {{Peters}, P.~C.},
        title = "{Gravitational Radiation and the Motion of Two Point Masses}",
      journal = {Physical Review},
         year = 1964,
        month = nov,
       volume = {136},
       number = {4B},
        pages = {1224-1232},
          doi = {10.1103/PhysRev.136.B1224},
       adsurl = {https://ui.adsabs.harvard.edu/abs/1964PhRv..136.1224P}
}

@ARTICLE{Samsing18a,
       author = {{Samsing}, Johan and {Ilan}, Teva},
        title = "{Topology of black hole binary-single interactions}",
      journal = {\mnras},
         year = 2018,
        month = may,
       volume = {476},
       number = {2},
        pages = {1548-1560},
          doi = {10.1093/mnras/sty197},
archivePrefix = {arXiv},
       eprint = {1706.04672},
 primaryClass = {astro-ph.HE},
       adsurl = {https://ui.adsabs.harvard.edu/abs/2018MNRAS.476.1548S}
}

@article{Lee:2010in,
author = {Lee, William H and Ramirez-Ruiz, Enrico and van de Ven, Glenn},
title = {{SHORT GAMMA-RAY BURSTS FROM DYNAMICALLY ASSEMBLED COMPACT BINARIES IN GLOBULAR CLUSTERS: PATHWAYS, RATES, HYDRODYNAMICS, AND COSMOLOGICAL SETTING}},
journal = {\apj},
year = {2010},
volume = {720},
number = {1},
pages = {953--975},
month = aug
}

@article{2000ApJ...528L..17P,
author = {Portegies Zwart, Simon F and McMillan, Stephen L W},
title = {{Black Hole Mergers in the Universe}},
journal = {\apj},
year = {2000},
volume = {528},
number = {1},
pages = {L17--L20},
month = jan
}

@article{2006ApJ...640..156G,
author = {G{\"u}ltekin, Kayhan and Miller, M Coleman and Hamilton, Douglas P},
title = {{Three-Body Dynamics with Gravitational Wave Emission}},
journal = {\apj},
year = {2006},
volume = {640},
number = {1},
pages = {156--166},
month = mar
}

@article{Hut:1983js,
author = {Hut, P and Bahcall, J.~N.},
title = {{Binary-single star scattering. I - Numerical experiments for equal masses}},
journal = {\apj},
year = {1983},
volume = {268},
pages = {319--341},
month = may
}

@ARTICLE{2024A&A...689A..24T,
       author = {{Trani}, Alessandro Alberto and {Leigh}, Nathan W.~C. and {Boekholt}, Tjarda C.~N. and {Portegies Zwart}, Simon},
        title = "{Isles of regularity in a sea of chaos amid the gravitational three-body problem}",
      journal = {\aap},
         year = 2024,
        month = sep,
       volume = {689},
          eid = {A24},
        pages = {A24},
          doi = {10.1051/0004-6361/202449862},
archivePrefix = {arXiv},
       eprint = {2403.03247},
 primaryClass = {astro-ph.EP},
       adsurl = {https://ui.adsabs.harvard.edu/abs/2024A&A...689A..24T}
}

@ARTICLE{2024A&A...683A.135T,
       author = {{Trani}, A.~A. and {Quaini}, S. and {Colpi}, M.},
        title = "{Three-body encounters in black hole discs around a supermassive black hole. The disc velocity dispersion and the Keplerian tidal field determine the eccentricity and spin-orbit alignment of gravitational wave mergers}",
      journal = {\aap},
         year = 2024,
        month = mar,
       volume = {683},
          eid = {A135},
        pages = {A135},
          doi = {10.1051/0004-6361/202347920},
archivePrefix = {arXiv},
       eprint = {2312.13281},
 primaryClass = {astro-ph.HE},
       adsurl = {https://ui.adsabs.harvard.edu/abs/2024A&A...683A.135T}
}

@ARTICLE{2022ApJ...926...83T,
       author = {{Tanikawa}, Ataru and {Yoshida}, Takashi and {Kinugawa}, Tomoya and {Trani}, Alessandro A. and {Hosokawa}, Takashi and {Susa}, Hajime and {Omukai}, Kazuyuki},
        title = "{Merger Rate Density of Binary Black Holes through Isolated Population I, II, III and Extremely Metal-poor Binary Star Evolution}",
      journal = {\apj},
         year = 2022,
        month = feb,
       volume = {926},
       number = {1},
          eid = {83},
        pages = {83},
          doi = {10.3847/1538-4357/ac4247},
archivePrefix = {arXiv},
       eprint = {2110.10846},
 primaryClass = {astro-ph.HE},
       adsurl = {https://ui.adsabs.harvard.edu/abs/2022ApJ...926...83T}
}

@ARTICLE{2021ApJ...910...30T,
       author = {{Tanikawa}, Ataru and {Susa}, Hajime and {Yoshida}, Takashi and {Trani}, Alessandro A. and {Kinugawa}, Tomoya},
        title = "{Merger Rate Density of Population III Binary Black Holes Below, Above, and in the Pair-instability Mass Gap}",
      journal = {\apj},
         year = 2021,
        month = mar,
       volume = {910},
       number = {1},
          eid = {30},
        pages = {30},
          doi = {10.3847/1538-4357/abe40d},
archivePrefix = {arXiv},
       eprint = {2008.01890},
 primaryClass = {astro-ph.HE},
       adsurl = {https://ui.adsabs.harvard.edu/abs/2021ApJ...910...30T}
}

@ARTICLE{2019MNRAS.485..889S,
       author = {{Spera}, Mario and {Mapelli}, Michela and {Giacobbo}, Nicola and {Trani}, Alessandro A. and {Bressan}, Alessandro and {Costa}, Guglielmo},
        title = "{Merging black hole binaries with the SEVN code}",
      journal = {\mnras},
         year = 2019,
        month = may,
       volume = {485},
       number = {1},
        pages = {889-907},
          doi = {10.1093/mnras/stz359},
archivePrefix = {arXiv},
       eprint = {1809.04605},
 primaryClass = {astro-ph.HE},
       adsurl = {https://ui.adsabs.harvard.edu/abs/2019MNRAS.485..889S}
}

@ARTICLE{2018LRR....21....3A,
       author = {{Abbott}, B.~P. and {Abbott}, R. and {Abbott}, T.~D. and {Abernathy}, M.~R. and {Acernese}, F. and {Ackley}, K. and {Adams}, C. and {Adams}, T. and {Addesso}, P. and {Adhikari}, R.~X. and {Adya}, V.~B. and {Affeldt}, C. and {Agathos}, M. and {Agatsuma}, K. and {Aggarwal}, N. and {Aguiar}, O.~D. and {Aiello}, L. and {Ain}, A. and {Ajith}, P. and {Akutsu}, T. and {Allen}, B. and {Allocca}, A. and {Altin}, P.~A. and {Ananyeva}, A. and {Anderson}, S.~B. and {Anderson}, W.~G. and {Ando}, M. and {Appert}, S. and {Arai}, K. and {Araya}, A. and {Araya}, M.~C. and {Areeda}, J.~S. and {Arnaud}, N. and {Arun}, K.~G. and {Asada}, H. and {Ascenzi}, S. and {Ashton}, G. and {Aso}, Y. and {Ast}, M. and {Aston}, S.~M. and {Astone}, P. and {Atsuta}, S. and {Aufmuth}, P. and {Aulbert}, C. and {Avila-Alvarez}, A. and {Awai}, K. and {Babak}, S. and {Bacon}, P. and {Bader}, M.~K.~M. and {Baiotti}, L. and {Baker}, P.~T. and {Baldaccini}, F. and {Ballardin}, G. and {Ballmer}, S.~W. and {Barayoga}, J.~C. and {Barclay}, S.~E. and {Barish}, B.~C. and {Barker}, D. and {Barone}, F. and {Barr}, B. and {Barsotti}, L. and {Barsuglia}, M. and {Barta}, D. and {Bartlett}, J. and {Barton}, M.~A. and {Bartos}, I. and {Bassiri}, R. and {Basti}, A. and {Batch}, J.~C. and {Baune}, C. and {Bavigadda}, V. and {Bazzan}, M. and {B{\'e}csy}, B. and {Beer}, C. and {Bejger}, M. and {Belahcene}, I. and {Belgin}, M. and {Bell}, A.~S. and {Berger}, B.~K. and {Bergmann}, G. and {Berry}, C.~P.~L. and {Bersanetti}, D. and {Bertolini}, A. and {Betzwieser}, J. and {Bhagwat}, S. and {Bhandare}, R. and {Bilenko}, I.~A. and {Billingsley}, G. and {Billman}, C.~R. and {Birch}, J. and {Birney}, R. and {Birnholtz}, O. and {Biscans}, S. and {Bisht}, A. and {Bitossi}, M. and {Biwer}, C. and {Bizouard}, M.~A. and {Blackburn}, J.~K. and {Blackman}, J. and {Blair}, C.~D. and {Blair}, D.~G. and {Blair}, R.~M. and {Bloemen}, S. and {Bock}, O. and {Boer}, M. and {Bogaert}, G. and {Bohe}, A. and {Bondu}, F. and {Bonnand}, R. and {Boom}, B.~A. and {Bork}, R. and {Boschi}, V. and {Bose}, S. and {Bouffanais}, Y. and {Bozzi}, A. and {Bradaschia}, C. and {Brady}, P.~R. and {Braginsky}, V.~B. and {Branchesi}, M. and {Brau}, J.~E. and {Briant}, T. and {Brillet}, A. and {Brinkmann}, M. and {Brisson}, V. and {Brockill}, P. and {Broida}, J.~E. and {Brooks}, A.~F. and {Brown}, D.~A. and {Brown}, D.~D. and {Brown}, N.~M. and {Brunett}, S. and {Buchanan}, C.~C. and {Buikema}, A. and {Bulik}, T. and {Bulten}, H.~J. and {Buonanno}, A. and {Buskulic}, D. and {Buy}, C. and {Byer}, R.~L. and {Cabero}, M. and {Cadonati}, L. and {Cagnoli}, G. and {Cahillane}, C. and {Calder{\'o}n Bustillo}, J. and {Callister}, T.~A. and {Calloni}, E. and {Camp}, J.~B. and {Cannon}, K.~C. and {Cao}, H. and {Cao}, J. and {Capano}, C.~D. and {Capocasa}, E. and {Carbognani}, F. and {Caride}, S. and {Casanueva Diaz}, J. and {Casentini}, C. and {Caudill}, S. and {Cavagli{\`a}}, M. and {Cavalier}, F. and {Cavalieri}, R. and {Cella}, G. and {Cepeda}, C.~B. and {Cerboni Baiardi}, L. and {Cerretani}, G. and {Cesarini}, E. and {Chamberlin}, S.~J. and {Chan}, M. and {Chao}, S. and {Charlton}, P. and {Chassande-Mottin}, E. and {Cheeseboro}, B.~D. and {Chen}, H.~Y. and {Chen}, Y. and {Cheng}, H. -P. and {Chincarini}, A. and {Chiummo}, A. and {Chmiel}, T. and {Cho}, H.~S. and {Cho}, M. and {Chow}, J.~H. and {Christensen}, N. and {Chu}, Q. and {Chua}, A.~J.~K. and {Chua}, S. and {Chung}, S. and {Ciani}, G. and {Clara}, F. and {Clark}, J.~A. and {Cleva}, F. and {Cocchieri}, C. and {Coccia}, E. and {Cohadon}, P. -F. and {Colla}, A. and {Collette}, C.~G. and {Cominsky}, L. and {Constancio}, M. and {Conti}, L. and {Cooper}, S.~J. and {Corbitt}, T.~R. and {Cornish}, N. and {Corsi}, A. and {Cortese}, S. and {Costa}, C.~A. and {Coughlin}, M.~W. and {Coughlin}, S.~B. and {Coulon}, J. -P. and {Countryman}, S.~T. and {Couvares}, P. and {Covas}, P.~B. and {Cowan}, E.~E. and {Coward}, D.~M. and {Cowart}, M.~J. and {Coyne}, D.~C. and {Coyne}, R. and {Creighton}, J.~D.~E. and {Creighton}, T.~D. and {Cripe}, J. and {Crowder}, S.~G. and {Cullen}, T.~J. and {Cumming}, A. and {Cunningham}, L. and {Cuoco}, E. and {Canton}, T. Dal and {Danilishin}, S.~L. and {D'Antonio}, S. and {Danzmann}, K. and {Dasgupta}, A. and {da Silva Costa}, C.~F. and {Dattilo}, V. and {Dave}, I. and {Davier}, M. and {Davies}, G.~S. and {Davis}, D. and {Daw}, E.~J. and {Day}, B. and {Day}, R. and {de}, S. and {Debra}, D. and {Debreczeni}, G. and {Degallaix}, J. and {de Laurentis}, M. and {Del{\'e}glise}, S. and {Del Pozzo}, W. and {Denker}, T. and {Dent}, T. and {Dergachev}, V. and {De Rosa}, R. and {Derosa}, R.~T. and {Desalvo}, R. and {Devine}, R.~C. and {Dhurandhar}, S. and {D{\'\i}az}, M.~C. and {Fiore}, L. Di and {Giovanni}, M. Di and {Girolamo}, T. Di and {Lieto}, A. Di and {Pace}, S. Di and {Palma}, I. Di and {Virgilio}, A. Di and {Doctor}, Z. and {Doi}, K. and {Dolique}, V. and {Donovan}, F. and {Dooley}, K.~L. and {Doravari}, S. and {Dorrington}, I. and {Douglas}, R. and {Dovale {\'A}lvarez}, M. and {Downes}, T.~P. and {Drago}, M. and {Drever}, R.~W.~P. and {Driggers}, J.~C. and {Du}, Z. and {Ducrot}, M. and {Dwyer}, S.~E. and {Eda}, K. and {Edo}, T.~B. and {Edwards}, M.~C. and {Effler}, A. and {Eggenstein}, H. -B. and {Ehrens}, P. and {Eichholz}, J. and {Eikenberry}, S.~S. and {Eisenstein}, R.~A. and {Essick}, R.~C. and {Etienne}, Z. and {Etzel}, T. and {Evans}, M. and {Evans}, T.~M. and {Everett}, R. and {Factourovich}, M. and {Fafone}, V. and {Fair}, H. and {Fairhurst}, S. and {Fan}, X. and {Farinon}, S. and {Farr}, B. and {Farr}, W.~M. and {Fauchon-Jones}, E.~J. and {Favata}, M. and {Fays}, M. and {Fehrmann}, H. and {Fejer}, M.~M. and {Fern{\'a}ndez Galiana}, A. and {Ferrante}, I. and {Ferreira}, E.~C. and {Ferrini}, F. and {Fidecaro}, F. and {Fiori}, I. and {Fiorucci}, D. and {Fisher}, R.~P. and {Flaminio}, R. and {Fletcher}, M. and {Fong}, H. and {Forsyth}, S.~S. and {Fournier}, J. -D. and {Frasca}, S. and {Frasconi}, F. and {Frei}, Z. and {Freise}, A. and {Frey}, R. and {Frey}, V. and {Fries}, E.~M. and {Fritschel}, P. and {Frolov}, V.~V. and {Fujii}, Y. and {Fujimoto}, M. -K. and {Fulda}, P. and {Fyffe}, M. and {Gabbard}, H. and {Gadre}, B.~U. and {Gaebel}, S.~M. and {Gair}, J.~R. and {Gammaitoni}, L. and {Gaonkar}, S.~G. and {Garufi}, F. and {Gaur}, G. and {Gayathri}, V. and {Gehrels}, N. and {Gemme}, G. and {Genin}, E. and {Gennai}, A. and {George}, J. and {Gergely}, L. and {Germain}, V. and {Ghonge}, S. and {Ghosh}, Abhirup and {Ghosh}, Archisman and {Ghosh}, S. and {Giaime}, J.~A. and {Giardina}, K.~D. and {Giazotto}, A. and {Gill}, K. and {Glaefke}, A. and {Goetz}, E. and {Goetz}, R. and {Gondan}, L. and {Gonz{\'a}lez}, G. and {Gonzalez Castro}, J.~M. and {Gopakumar}, A. and {Gorodetsky}, M.~L. and {Gossan}, S.~E. and {Gosselin}, M. and {Gouaty}, R. and {Grado}, A. and {Graef}, C. and {Granata}, M. and {Grant}, A. and {Gras}, S. and {Gray}, C. and {Greco}, G. and {Green}, A.~C. and {Groot}, P. and {Grote}, H. and {Grunewald}, S. and {Guidi}, G.~M. and {Guo}, X. and {Gupta}, A. and {Gupta}, M.~K. and {Gushwa}, K.~E. and {Gustafson}, E.~K. and {Gustafson}, R. and {Hacker}, J.~J. and {Hagiwara}, A. and {Hall}, B.~R. and {Hall}, E.~D. and {Hammond}, G. and {Haney}, M. and {Hanke}, M.~M. and {Hanks}, J. and {Hanna}, C. and {Hannam}, M.~D. and {Hanson}, J. and {Hardwick}, T. and {Harms}, J. and {Harry}, G.~M. and {Harry}, I.~W. and {Hart}, M.~J. and {Hartman}, M.~T. and {Haster}, C. -J. and {Haughian}, K. and {Hayama}, K. and {Healy}, J. and {Heidmann}, A. and {Heintze}, M.~C. and {Heitmann}, H. and {Hello}, P. and {Hemming}, G. and {Hendry}, M. and {Heng}, I.~S. and {Hennig}, J. and {Henry}, J. and {Heptonstall}, A.~W. and {Heurs}, M. and {Hild}, S. and {Hirose}, E. and {Hoak}, D. and {Hofman}, D. and {Holt}, K. and {Holz}, D.~E. and {Hopkins}, P. and {Hough}, J. and {Houston}, E.~A. and {Howell}, E.~J. and {Hu}, Y.~M. and {Huerta}, E.~A. and {Huet}, D. and {Hughey}, B. and {Husa}, S. and {Huttner}, S.~H. and {Huynh-Dinh}, T. and {Indik}, N. and {Ingram}, D.~R. and {Inta}, R. and {Ioka}, K. and {Isa}, H.~N. and {Isac}, J. -M. and {Isi}, M. and {Isogai}, T. and {Itoh}, Y. and {Iyer}, B.~R. and {Izumi}, K. and {Jacqmin}, T. and {Jani}, K. and {Jaranowski}, P. and {Jawahar}, S. and {Jim{\'e}nez-Forteza}, F. and {Johnson}, W.~W. and {Jones}, D.~I. and {Jones}, R. and {Jonker}, R.~J.~G. and {Ju}, L. and {Junker}, J. and {Kagawa}, T. and {Kajita}, T. and {Kakizaki}, M. and {Kalaghatgi}, C.~V. and {Kalogera}, V. and {Kamiizumi}, M. and {Kanda}, N. and {Kandhasamy}, S. and {Kanemura}, S. and {Kaneyama}, M. and {Kang}, G. and {Kanner}, J.~B. and {Karki}, S. and {Karvinen}, K.~S. and {Kasprzack}, M. and {Kataoka}, Y. and {Katsavounidis}, E. and {Katzman}, W. and {Kaufer}, S. and {Kaur}, T. and {Kawabe}, K. and {Kawai}, N. and {Kawamura}, S. and {K{\'e}f{\'e}lian}, F. and {Keitel}, D. and {Kelley}, D.~B. and {Kennedy}, R. and {Key}, J.~S. and {Khalili}, F.~Y. and {Khan}, I. and {Khan}, S. and {Khan}, Z. and {Khazanov}, E.~A. and {Kijbunchoo}, N. and {Kim}, C. and {Kim}, H. and {Kim}, J.~C. and {Kim}, J. and {Kim}, W. and {Kim}, Y. -M. and {Kimbrell}, S.~J. and {Kimura}, N. and {King}, E.~J. and {King}, P.~J. and {Kirchhoff}, R. and {Kissel}, J.~S. and {Klein}, B. and {Kleybolte}, L. and {Klimenko}, S. and {Koch}, P. and {Koehlenbeck}, S.~M. and {Kojima}, Y. and {Kokeyama}, K. and {Koley}, S. and {Komori}, K. and {Kondrashov}, V. and {Kontos}, A. and {Korobko}, M. and {Korth}, W.~Z. and {Kotake}, K. and {Kowalska}, I. and {Kozak}, D.~B. and {Kr{\"a}mer}, C. and {Kringel}, V. and {Krishnan}, B. and {Kr{\'o}lak}, A. and {Kuehn}, G. and {Kumar}, P. and {Kumar}, Rahul and {Kumar}, Rakesh and {Kuo}, L. and {Kuroda}, K. and {Kutynia}, A. and {Kuwahara}, Y. and {Lackey}, B.~D. and {Landry}, M. and {Lang}, R.~N. and {Lange}, J. and {Lantz}, B. and {Lanza}, R.~K. and {Lartaux-Vollard}, A. and {Lasky}, P.~D. and {Laxen}, M. and {Lazzarini}, A. and {Lazzaro}, C. and {Leaci}, P. and {Leavey}, S. and {Lebigot}, E.~O. and {Lee}, C.~H. and {Lee}, H.~K. and {Lee}, H.~M. and {Lee}, H.~W. and {Lee}, K. and {Lehmann}, J. and {Lenon}, A. and {Leonardi}, M. and {Leong}, J.~R. and {Leroy}, N. and {Letendre}, N. and {Levin}, Y. and {Li}, T.~G.~F. and {Libson}, A. and {Littenberg}, T.~B. and {Liu}, J. and {Lockerbie}, N.~A. and {Lombardi}, A.~L. and {London}, L.~T. and {Lord}, J.~E. and {Lorenzini}, M. and {Loriette}, V. and {Lormand}, M. and {Losurdo}, G. and {Lough}, J.~D. and {Lousto}, C.~O. and {Lovelace}, G. and {L{\"u}ck}, H. and {Lundgren}, A.~P. and {Lynch}, R. and {Ma}, Y. and {Macfoy}, S. and {Machenschalk}, B. and {Macinnis}, M. and {MacLeod}, D.~M. and {Maga{\~n}a-Sandoval}, F. and {Majorana}, E. and {Maksimovic}, I. and {Malvezzi}, V. and {Man}, N. and {Mandic}, V. and {Mangano}, V. and {Mano}, S. and {Mansell}, G.~L. and {Manske}, M. and {Mantovani}, M. and {Marchesoni}, F. and {Marchio}, M. and {Marion}, F. and {M{\'a}rka}, S. and {M{\'a}rka}, Z. and {Markosyan}, A.~S. and {Maros}, E. and {Martelli}, F. and {Martellini}, L. and {Martin}, I.~W. and {Martynov}, D.~V. and {Mason}, K. and {Masserot}, A. and {Massinger}, T.~J. and {Masso-Reid}, M. and {Mastrogiovanni}, S. and {Matichard}, F. and {Matone}, L. and {Matsumoto}, N. and {Matsushima}, F. and {Mavalvala}, N. and {Mazumder}, N. and {McCarthy}, R. and {McClelland}, D.~E. and {McCormick}, S. and {McGrath}, C. and {McGuire}, S.~C. and {McIntyre}, G. and {McIver}, J. and {McManus}, D.~J. and {McRae}, T. and {McWilliams}, S.~T. and {Meacher}, D. and {Meadors}, G.~D. and {Meidam}, J. and {Melatos}, A. and {Mendell}, G. and {Mendoza-Gandara}, D. and {Mercer}, R.~A. and {Merilh}, E.~L. and {Merzougui}, M. and {Meshkov}, S. and {Messenger}, C. and {Messick}, C. and {Metzdorff}, R. and {Meyers}, P.~M. and {Mezzani}, F. and {Miao}, H. and {Michel}, C. and {Michimura}, Y. and {Middleton}, H. and {Mikhailov}, E.~E. and {Milano}, L. and {Miller}, A.~L. and {Miller}, A. and {Miller}, B.~B. and {Miller}, J. and {Millhouse}, M. and {Minenkov}, Y. and {Ming}, J. and {Mirshekari}, S. and {Mishra}, C. and {Mitrofanov}, V.~P. and {Mitselmakher}, G. and {Mittleman}, R. and {Miyakawa}, O. and {Miyamoto}, A. and {Miyamoto}, T. and {Miyoki}, S. and {Moggi}, A. and {Mohan}, M. and {Mohapatra}, S.~R.~P. and {Montani}, M. and {Moore}, B.~C. and {Moore}, C.~J. and {Moraru}, D. and {Moreno}, G. and {Morii}, W. and {Morisaki}, S. and {Moriwaki}, Y. and {Morriss}, S.~R. and {Mours}, B. and {Mow-Lowry}, C.~M. and {Mueller}, G. and {Muir}, A.~W. and {Mukherjee}, Arunava and {Mukherjee}, D. and {Mukherjee}, S. and {Mukund}, N. and {Mullavey}, A. and {Munch}, J. and {Muniz}, E.~A.~M. and {Murray}, P.~G. and {Mytidis}, A. and {Nagano}, S. and {Nakamura}, K. and {Nakamura}, T. and {Nakano}, H. and {Nakano}, Masaya and {Nakano}, Masayuki and {Nakao}, K. and {Napier}, K. and {Nardecchia}, I. and {Narikawa}, T. and {Naticchioni}, L. and {Nelemans}, G. and {Nelson}, T.~J.~N. and {Neri}, M. and {Nery}, M. and {Neunzert}, A. and {Newport}, J.~M. and {Newton}, G. and {Nguyen}, T.~T. and {Ni}, W. -T. and {Nielsen}, A.~B. and {Nissanke}, S. and {Nitz}, A. and {Noack}, A. and {Nocera}, F. and {Nolting}, D. and {Normandin}, M.~E.~N. and {Nuttall}, L.~K. and {Oberling}, J. and {Ochsner}, E. and {Oelker}, E. and {Ogin}, G.~H. and {Oh}, J.~J. and {Oh}, S.~H. and {Ohashi}, M. and {Ohishi}, N. and {Ohkawa}, M. and {Ohme}, F. and {Okutomi}, K. and {Oliver}, M. and {Ono}, K. and {Ono}, Y. and {Oohara}, K. and {Oppermann}, P. and {Oram}, Richard J. and {O'Reilly}, B. and {O'Shaughnessy}, R. and {Ottaway}, D.~J. and {Overmier}, H. and {Owen}, B.~J. and {Pace}, A.~E. and {Page}, J. and {Pai}, A. and {Pai}, S.~A. and {Palamos}, J.~R. and {Palashov}, O. and {Palomba}, C. and {Pal-Singh}, A. and {Pan}, H. and {Pankow}, C. and {Pannarale}, F. and {Pant}, B.~C. and {Paoletti}, F. and {Paoli}, A. and {Papa}, M.~A. and {Paris}, H.~R. and {Parker}, W. and {Pascucci}, D. and {Pasqualetti}, A. and {Passaquieti}, R. and {Passuello}, D. and {Patricelli}, B. and {Pearlstone}, B.~L. and {Pedraza}, M. and {Pedurand}, R. and {Pekowsky}, L. and {Pele}, A. and {Pe{\~n}a Arellano}, F.~E. and {Penn}, S. and {Perez}, C.~J. and {Perreca}, A. and {Perri}, L.~M. and {Pfeiffer}, H.~P. and {Phelps}, M. and {Piccinni}, O.~J. and {Pichot}, M. and {Piergiovanni}, F. and {Pierro}, V. and {Pillant}, G. and {Pinard}, L. and {Pinto}, I.~M. and {Pitkin}, M. and {Poe}, M. and {Poggiani}, R. and {Popolizio}, P. and {Post}, A. and {Powell}, J. and {Prasad}, J. and {Pratt}, J.~W.~W. and {Predoi}, V. and {Prestegard}, T. and {Prijatelj}, M. and {Principe}, M. and {Privitera}, S. and {Prodi}, G.~A. and {Prokhorov}, L.~G. and {Puncken}, O. and {Punturo}, M. and {Puppo}, P. and {P{\"u}rrer}, M. and {Qi}, H. and {Qin}, J. and {Qiu}, S. and {Quetschke}, V. and {Quintero}, E.~A. and {Quitzow-James}, R. and {Raab}, F.~J. and {Rabeling}, D.~S. and {Radkins}, H. and {Raffai}, P. and {Raja}, S. and {Rajan}, C. and {Rakhmanov}, M. and {Rapagnani}, P. and {Raymond}, V. and {Razzano}, M. and {Re}, V. and {Read}, J. and {Regimbau}, T. and {Rei}, L. and {Reid}, S. and {Reitze}, D.~H. and {Rew}, H. and {Reyes}, S.~D. and {Rhoades}, E. and {Ricci}, F. and {Riles}, K. and {Rizzo}, M. and {Robertson}, N.~A. and {Robie}, R. and {Robinet}, F. and {Rocchi}, A. and {Rolland}, L. and {Rollins}, J.~G. and {Roma}, V.~J. and {Romano}, R. and {Romie}, J.~H. and {Rosi{\'n}ska}, D. and {Rowan}, S. and {R{\"u}diger}, A. and {Ruggi}, P. and {Ryan}, K. and {Sachdev}, S. and {Sadecki}, T. and {Sadeghian}, L. and {Sago}, N. and {Saijo}, M. and {Saito}, Y. and {Sakai}, K. and {Sakellariadou}, M. and {Salconi}, L. and {Saleem}, M. and {Salemi}, F. and {Samajdar}, A. and {Sammut}, L. and {Sampson}, L.~M. and {Sanchez}, E.~J. and {Sandberg}, V. and {Sanders}, J.~R. and {Sasaki}, Y. and {Sassolas}, B. and {Sathyaprakash}, B.~S. and {Sato}, S. and {Sato}, T. and {Saulson}, P.~R. and {Sauter}, O. and {Savage}, R.~L. and {Sawadsky}, A. and {Schale}, P. and {Scheuer}, J. and {Schmidt}, E. and {Schmidt}, J. and {Schmidt}, P. and {Schnabel}, R. and {Schofield}, R.~M.~S. and {Sch{\"o}nbeck}, A. and {Schreiber}, E. and {Schuette}, D. and {Schutz}, B.~F. and {Schwalbe}, S.~G. and {Scott}, J. and {Scott}, S.~M. and {Sekiguchi}, T. and {Sekiguchi}, Y. and {Sellers}, D. and {Sengupta}, A.~S. and {Sentenac}, D. and {Sequino}, V. and {Sergeev}, A. and {Setyawati}, Y. and {Shaddock}, D.~A. and {Shaffer}, T.~J. and {Shahriar}, M.~S. and {Shapiro}, B. and {Shawhan}, P. and {Sheperd}, A. and {Shibata}, M. and {Shikano}, Y. and {Shimoda}, T. and {Shoda}, A. and {Shoemaker}, D.~H. and {Shoemaker}, D.~M. and {Siellez}, K. and {Siemens}, X. and {Sieniawska}, M. and {Sigg}, D. and {Silva}, A.~D. and {Singer}, A. and {Singer}, L.~P. and {Singh}, A. and {Singh}, R. and {Singhal}, A. and {Sintes}, A.~M. and {Slagmolen}, B.~J.~J. and {Smith}, B. and {Smith}, J.~R. and {Smith}, R.~J.~E. and {Somiya}, K. and {Son}, E.~J. and {Sorazu}, B. and {Sorrentino}, F. and {Souradeep}, T. and {Spencer}, A.~P. and {Srivastava}, A.~K. and {Staley}, A. and {Steinke}, M. and {Steinlechner}, J. and {Steinlechner}, S. and {Steinmeyer}, D. and {Stephens}, B.~C. and {Stevenson}, S.~P. and {Stone}, R. and {Strain}, K.~A. and {Straniero}, N. and {Stratta}, G. and {Strigin}, S.~E. and {Sturani}, R. and {Stuver}, A.~L. and {Sugimoto}, Y. and {Summerscales}, T.~Z. and {Sun}, L. and {Sunil}, S. and {Sutton}, P.~J. and {Suzuki}, T. and {Swinkels}, B.~L. and {Szczepa{\'n}czyk}, M.~J. and {Tacca}, M. and {Tagoshi}, H. and {Takada}, S. and {Takahashi}, H. and {Takahashi}, R. and {Takamori}, A. and {Talukder}, D. and {Tanaka}, H. and {Tanaka}, K. and {Tanaka}, T. and {Tanner}, D.~B. and {T{\'a}pai}, M. and {Taracchini}, A. and {Tatsumi}, D. and {Taylor}, R. and {Telada}, S. and {Theeg}, T. and {Thomas}, E.~G. and {Thomas}, M. and {Thomas}, P. and {Thorne}, K.~A. and {Thrane}, E. and {Tippens}, T. and {Tiwari}, S. and {Tiwari}, V. and {Tokmakov}, K.~V. and {Toland}, K. and {Tomaru}, T. and {Tomlinson}, C. and {Tonelli}, M. and {Tornasi}, Z. and {Torrie}, C.~I. and {T{\"o}yr{\"a}}, D. and {Travasso}, F. and {Traylor}, G. and {Trifir{\`o}}, D. and {Trinastic}, J. and {Tringali}, M.~C. and {Trozzo}, L. and {Tse}, M. and {Tso}, R. and {Tsubono}, K. and {Tsuzuki}, T. and {Turconi}, M. and {Tuyenbayev}, D. and {Uchiyama}, T. and {Uehara}, T. and {Ueki}, S. and {Ueno}, K. and {Ugolini}, D. and {Unnikrishnan}, C.~S. and {Urban}, A.~L. and {Ushiba}, T. and {Usman}, S.~A. and {Vahlbruch}, H. and {Vajente}, G. and {Valdes}, G. and {van Bakel}, N. and {van Beuzekom}, M. and {van den Brand}, J.~F.~J. and {van den Broeck}, C. and {Vander-Hyde}, D.~C. and {van der Schaaf}, L. and {van Heijningen}, J.~V. and {van Putten}, M.~H.~P.~M. and {van Veggel}, A.~A. and {Vardaro}, M. and {Varma}, V. and {Vass}, S. and {Vas{\'u}th}, M. and {Vecchio}, A. and {Vedovato}, G. and {Veitch}, J. and {Veitch}, P.~J. and {Venkateswara}, K. and {Venugopalan}, G. and {Verkindt}, D. and {Vetrano}, F. and {Vicer{\'e}}, A. and {Viets}, A.~D. and {Vinciguerra}, S. and {Vine}, D.~J. and {Vinet}, J. -Y. and {Vitale}, S. and {Vo}, T. and {Vocca}, H. and {Vorvick}, C. and {Voss}, D.~V. and {Vousden}, W.~D. and {Vyatchanin}, S.~P. and {Wade}, A.~R. and {Wade}, L.~E. and {Wade}, M. and {Wakamatsu}, T. and {Walker}, M. and {Wallace}, L. and {Walsh}, S. and {Wang}, G. and {Wang}, H. and {Wang}, M. and {Wang}, Y. and {Ward}, R.~L. and {Warner}, J. and {Was}, M. and {Watchi}, J. and {Weaver}, B. and {Wei}, L. -W. and {Weinert}, M. and {Weinstein}, A.~J. and {Weiss}, R. and {Wen}, L. and {We{\ss}els}, P. and {Westphal}, T. and {Wette}, K. and {Whelan}, J.~T. and {Whiting}, B.~F. and {Whittle}, C. and {Williams}, D. and {Williams}, R.~D. and {Williamson}, A.~R. and {Willis}, J.~L. and {Willke}, B. and {Wimmer}, M.~H. and {Winkler}, W. and {Wipf}, C.~C. and {Wittel}, H. and {Woan}, G. and {Woehler}, J. and {Worden}, J. and {Wright}, J.~L. and {Wu}, D.~S. and {Wu}, G. and {Yam}, W. and {Yamamoto}, H. and {Yamamoto}, K. and {Yamamoto}, T. and {Yancey}, C.~C. and {Yano}, K. and {Yap}, M.~J. and {Yokoyama}, J. and {Yokozawa}, T. and {Yoon}, T.~H. and {Yu}, Hang and {Yu}, Haocun and {Yuzurihara}, H. and {Yvert}, M. and {Zadro{\.z}ny}, A. and {Zangrando}, L. and {Zanolin}, M. and {Zeidler}, S. and {Zendri}, J. -P. and {Zevin}, M. and {Zhang}, L. and {Zhang}, M. and {Zhang}, T. and {Zhang}, Y. and {Zhao}, C. and {Zhou}, M. and {Zhou}, Z. and {Zhu}, S.~J. and {Zhu}, X.~J. and {Zucker}, M.~E. and {Zweizig}, J. and {Kagra Collaboration}, Ligo Scientific Collaboration and {VIRGO Collaboration}},
        title = "{Prospects for observing and localizing gravitational-wave transients with Advanced LIGO, Advanced Virgo and KAGRA}",
      journal = {Living Reviews in Relativity},
         year = 2018,
        month = dec,
       volume = {21},
       number = {1},
          eid = {3},
        pages = {3},
          doi = {10.1007/s41114-018-0012-9},
archivePrefix = {arXiv},
       eprint = {1304.0670},
 primaryClass = {gr-qc},
       adsurl = {https://ui.adsabs.harvard.edu/abs/2018LRR....21....3A}
}

@ARTICLE{2014ApJ...784...71S,
   author = {{Samsing}, J. and {MacLeod}, M. and {Ramirez-Ruiz}, E.},
    title = "{The Formation of Eccentric Compact Binary Inspirals and the Role of Gravitational Wave Emission in Binary-Single Stellar Encounters}",
  journal = {\apj},
archivePrefix = "arXiv",
   eprint = {1308.2964},
 primaryClass = "astro-ph.HE",
     year = 2014,
    month = mar,
   volume = 784,
      eid = {71},
    pages = {71},
      doi = {10.1088/0004-637X/784/1/71},
   adsurl = {http://adsabs.harvard.edu/abs/2014ApJ...784...71S}
}

@ARTICLE{2016PhRvD..93h4029R,
   author = {{Rodriguez}, C.~L. and {Chatterjee}, S. and {Rasio}, F.~A.},
    title = "{Binary black hole mergers from globular clusters: Masses, merger rates, and the impact of stellar evolution}",
  journal = {\prd},
archivePrefix = "arXiv",
   eprint = {1602.02444},
 primaryClass = "astro-ph.HE",
     year = 2016,
    month = apr,
   volume = 93,
   number = 8,
      eid = {084029},
    pages = {084029},
      doi = {10.1103/PhysRevD.93.084029},
   adsurl = {http://adsabs.harvard.edu/abs/2016PhRvD..93h4029R}
}

@ARTICLE{2016ApJ...824L...8R,
   author = {{Rodriguez}, C.~L. and {Haster}, C.-J. and {Chatterjee}, S. and 
	{Kalogera}, V. and {Rasio}, F.~A.},
    title = "{Dynamical Formation of the GW150914 Binary Black Hole}",
  journal = {\apjl},
archivePrefix = "arXiv",
   eprint = {1604.04254},
 primaryClass = "astro-ph.HE",
     year = 2016,
    month = jun,
   volume = 824,
      eid = {L8},
    pages = {L8},
      doi = {10.3847/2041-8205/824/1/L8},
   adsurl = {http://adsabs.harvard.edu/abs/2016ApJ...824L...8R}
}

@ARTICLE{2015ApJ...802L..22R,
   author = {{Ramirez-Ruiz}, E. and {Trenti}, M. and {MacLeod}, M. and {Roberts}, L.~F. and 
	{Lee}, W.~H. and {Saladino-Rosas}, M.~I.},
    title = "{Compact Stellar Binary Assembly in the First Nuclear Star Clusters and r-process Synthesis in the Early Universe}",
  journal = {\apjl},
archivePrefix = "arXiv",
   eprint = {1410.3467},
     year = 2015,
    month = apr,
   volume = 802,
      eid = {L22},
    pages = {L22},
      doi = {10.1088/2041-8205/802/2/L22},
   adsurl = {http://adsabs.harvard.edu/abs/2015ApJ...802L..22R}
}

\newpage

\end{document}